\documentclass[12pt]{article}
\usepackage{amsmath}
\usepackage{amssymb}
\usepackage{graphicx}
\usepackage[numbers,sort&compress]{natbib}
\title{Inverse scattering transform for the Tzitz\'{e}ica equation}
\author{Linlin Wang$^a$\thanks{wanglinlinzzu@sina.com}, Junyi Zhu$^b$
\\
{\small $^a$ College of Science, Henan University of  Engineering, }\\
{\small Zhengzhou 451191, Henan, China}\\
{\small $^b$
School of Mathematics and Statistics, Zhengzhou University,}\\
{\small Zhengzhou 450001, Henan, China}\\
}
\date{}
\begin{document}
\maketitle
\begin{abstract}
The inverse scattering transform is extended to investigate the
Tzitz\'{e}ica equation. A set of sectionally analytic eigenfunctions
and auxiliary eigenfunctions are introduced. We note that in this
procedure, the auxiliary eigenfunctions play an important role.
Besides, the symmetries of the analytic eigenfunctions and scattering
data are discussed. The asymptotic behaviors of the Jost eigenfunctions are derived
systematically. A Riemann-Hilbert problem is constructed to study
the inverse scattering problem. Lastly, some novel exact solutions
are obtained for reflectionless potentials.

\end{abstract}
\section{Introduction}
The inverse scattering transform method is an effective tool to
study the integrable nonlinear equation with sufficient decay boundary condition or non-zero boundary condition. Besides, some novel and interesting properties for solutions can be given by the inverse scattering transform. There are so many works involving these problems. We here only refer to some of them \cite{prl19-1095,A-C1991,jmp47-063508,non29-915,jmp59-011501,cpam60-951,pd399-173,pd402-132170,cnsns67-555,jmaa474-452}.
In contrast to $2\times2$ matrix linear spectral problems, the $3\times3$ or higher order matrix problems are more difficult to be investigated by the inverse scattering transform method \cite{sam55-9,cpam37-39,pd6-51,non23-2559}.

In the present paper, we will study the Tzitz\'{e}ica equation by the inverse scattering transform
\begin{equation}\label{a1}
u_{xt}=e^{u}-e^{-2u},
\end{equation}
where $u=u(x,t)$ and has sufficient decay for all $t$ as
$|x|\to\infty$. Here, the subscripted variables $x$ and $t$ in Eq.
(\ref{a1}) denote the corresponding partial differentiation. We note
that the Tzitz\'{e}ica equation (\ref{a1}) can be rewritten as other
forms \cite{amc177-745,mcm52-1834}
\begin{equation}\label{a2}
(\ln h)_{xt}=h-h^{-2},
\end{equation}
\begin{equation}\label{a3}
u_{xx}-u_{tt}=e^{u}-e^{-2u},
\end{equation}
 It can be directly calculated that Eq.(\ref{a1}) is equivalent to Eq. (\ref{a2}) with the
 transform $\ln h=u$. The Tzitz\'{e}ica equation (\ref{a1}) was initially
 proposed as the model describing special surfaces in differential geometry,
 where the ratio $k/d^{4}$ is constant. The parameter $k$ is the Gauss
 curvature of the surface and $d$ is the distance from the origin to the
 tangent plane at a fixed point.
In \cite{R-S2002,cmp77-21,pd3D-73}, the Tzitz\'{e}ica equation can
be reduced from the two-dimensional Toda lattice equation via the
conjugate nets. Furthermore, the relations between soliton theory
and affine differential geometry were presented in \cite{R-S2002}.

It is shown that Eq. (\ref{a1}) plays an
 important role in many fields, such as geometry \cite{mn10-331}, soliton
 theory \cite{prsa352-481,pd3D-73,ip10-711,dan247-1103} and gas dynamics \cite{pd26D-123}.
 The Tzitz\'{e}ica equation was studied by many methods,
 such as the Darboux transformation \cite{ctp45-577,pla211-94}, B\"{a}cklund
 transformation \cite{jmp40-2092}, the algebro-geometric
 approach \cite{imrp2015-2141,tmp82-111,ijmpa5-3021}, the Hirota bilinear method \cite{jgsp37-1,ijnsn10-935},
 the inverse scattering method \cite{pd3D-73,jmp34-5801} and the dressing
 method \cite{jgsp37-1}. In addition, there are some other
 results for the Tzitz\'{e}ica equation, such as the discretization \cite{pla223-55}, ultradiscretization
\cite{gmj47A-77}, symmetries \cite{cmp81-189}, conservation laws
\cite{prsa352-481} and others
\cite{gmj47A-221,jmp50-043511,jmp47-053502,jmp43-651,pd237-1079,cpl23-2885}.

This paper is to obtain the explicit solution of the Tzitz\'{e}ica
equation through inverse scattering transforms. To this end, a Riemann-Hilbert (RH) problem with the corresponding six jump matrices and six jump conditions is constructed. We note that it is difficult to construct a set of analyticity eigenfunctions in every domains. Thus, introducing the auxiliary eigenfunctions and the adjoint problem, we obtain the piecewise meromorphic function. To avoid the overly cumbersome work, we consider the symmetry conditions associated with the Tzitz\'{e}ica equation. By using the Cauchy projector, the RH problem is transformed to a system of matrix algebraic-integral equations. Based on the algebraic-integral equations, some explicit
solutions of Eq.(1.1) for the case of reflectionless potentials are given.

The paper is organized as follows. In section 2, the direct problem
of the Tzitz\'{e}ica equation Eq. (\ref{a1}) is considered. In
section 3, the discrete spectrum is discussed in detail. In section
4, the inverse problem is investigated. In the last section,  based on the
RH problem, some exact solutions of Eq. (\ref{a1}) are constructed.
\setcounter{equation}{0}
\section{Spectral analysis}
In this section, we will consider the direct scattering problem
associated with Eq. (\ref{a1}).
\subsection{Preliminaries: Lax pair, scattering matrix}
The Tzitz\'{e}ica equation (\ref{a1}) can be derived through the
compatibility condition of the following Lax pair
\begin{subequations}
\begin{equation}\label{b1a}
\psi_{0x}=U_{0}(x,\lambda,u) \psi_{0},
\end{equation}
\begin{equation}\label{b1b}
\psi_{0t}=V_{0}(t,\lambda,u)\psi_{0},
\end{equation}
\end{subequations}
where
\begin{equation}\label{b2}
U_{0}(x,\lambda,u)=\left(\begin{matrix}
0&0&\lambda e^{u}\\
e^{-u}&u_{x}&0\\
0&1&0
\end{matrix}\right),\quad
V_{0}(t,\lambda,u)=\left(\begin{matrix}
u_{t}&e^{-u}&0\\
0&0&e^{u}\\
\lambda^{-1}&0&0
\end{matrix}\right).
\end{equation}
Since the trace of the matrices $U_0$ and $V_0$ are not zero, we
make the following eigenfunction transformation
\begin{equation}\label{b3}
\left(\begin{array}{c}
\psi_{1}\\
\psi_{2}\\
\psi_{3}
\end{array}\right)=\left(\begin{matrix}
e^{-u}&0&0\\
0&\nu&0\\
0&0&\nu^{2}
\end{matrix}\right)
\left(\begin{array}{c}
\psi_{10}\\
\psi_{20}\\
\psi_{30}
\end{array}\right),
\end{equation}
and have the second Lax pair:
\begin{subequations}
\begin{equation}\label{b4a}
\psi_{x}=U(x,\lambda,u) \psi,
\end{equation}
\begin{equation}\label{b4b}
\psi_t=V(t,\lambda,u)\psi,
\end{equation}
\end{subequations}
with
\begin{equation}\label{b5}
U(x,\lambda,u)=\left(\begin{matrix}
-u_{x}&0&\nu\\
\nu&u_{x}&0\\
0&\nu&0
\end{matrix}\right),\quad
V(t,\lambda,u)=\left(\begin{matrix}
0&\frac{e^{-2u}}{\nu}&0\\
0&0&\frac{e^{u}}{\nu}\\
\frac{e^{u}}{\nu}&0&0
\end{matrix}\right),
\end{equation}
$\nu$ is the spectral parameter. Thus, the determinant of the
fundamental matrix solution of $(2.4)$ is $x$ and $t$ independent.
In the following, we will discuss the spectral analysis from the
spectral problem (2.4). Under the vanishing boundary condition
$u{\to} 0,x{\to}\pm\infty$, we have
\begin{equation}\label{b6}
\begin{aligned}
&U\to U_{\pm}=\left(\begin{matrix}
0&0&\nu\\
\nu&0&0\\
0&\nu&0
\end{matrix}\right), \quad x\to\pm\infty,
\\
&V\to V_{\pm}=\left(\begin{matrix}
0&\frac{1}{\nu}&0\\
0&0&\frac{1}{\nu}\\
\frac{1}{\nu}&0&0
\end{matrix}\right), \quad x\to\pm\infty.
\end{aligned}
\end{equation}
For convenience, we introduce an invertible matrix $A$, such that
\begin{equation}\label{b7}
A^{-1}U_{\pm}A=\nu\sigma, \quad A^{-1}V_{\pm}A=\nu^{-1}\sigma^{-1},
\end{equation}
where
\begin{eqnarray}\label{b8}
\sigma=\left(\begin{matrix}
1&0&0\\
0&\alpha&0\\
0&0&\alpha^{2}
\end{matrix}\right),\quad \quad
A=\left(\begin{matrix}
1&\alpha^{2}&\alpha\\
1&\alpha&\alpha^{2}\\
1&1&1
\end{matrix}\right).
\end{eqnarray}
Here, $\lambda=\nu^{3}, \alpha=e^{\frac{2}{3}\pi i}$. Now we
introduce the matrix solutions $\psi_{\pm}(x,t,\nu)$ of (\ref{b4a})
satisfying the following condition
\begin{equation}\label{b9}
\begin{aligned}
\psi_{\pm}(x,t,\nu)=Ae^{\nu x \sigma+\nu^{-1}\sigma^{-1}t}+o(1), \quad x\to\pm\infty.
\end{aligned}
\end{equation}
Thus
\begin{equation}\label{b10}
\det\psi_{\pm}(x,t,\nu)=3(\alpha-\alpha^{2})=\gamma.
\end{equation}
For the sake of analyticity, we introduce new Jost solutions
$\mu_{\pm}(x,t,\nu)$ defined by
\begin{equation}\label{b11}
\psi_{\pm}(x,t,\nu)=\mu_{\pm}(x,t,\nu)e^{\nu\sigma x+\nu^{-1}\sigma^{-1}t},
\end{equation}
with their corresponding asymptotic behaviour
\begin{equation}\label{b12}
\mu_\pm(x,t,\nu)\to A, \quad x\to\pm\infty.
\end{equation}
\vspace{1cm}
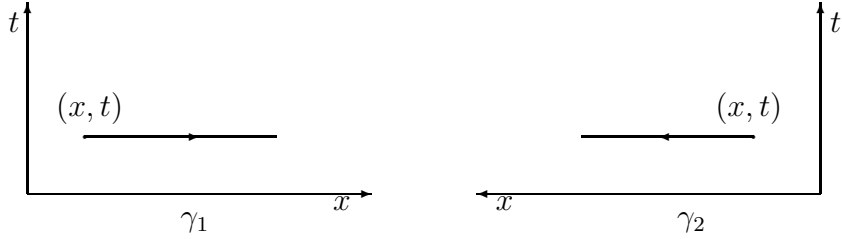
\begin{figure}[h]
\setlength{\unitlength}{0.1in}
\begin{picture}(10,10)
\put(9,3){\line(1,0){18}}
\put(26,3){\vector(1,0){1}}\put(25,2){\makebox(2,1)[l]{$x$}}
\put(9,3){\line(0,1){10}}
\put(9,12){\vector(0,1){1}}\put(8,11){\makebox(1,2)[l]{$t$}}
\put(12,6){\line(1,0){10}}\put(12,6){\circle{0}}
\put(17,6){\vector(1,0){1}}\put(10.5,6.5){\makebox(1,2)[l]{$(x,t)$}}
\put(17,1){\makebox(2, 1)[l]{$\gamma_{1}$}}
\put(32.5,3){\line(1,0){18}}
\put(33.5,3){\vector(-1,0){1}}\put(33.5,2){\makebox(2,1)[l]{$x$}}
\put(50.5,3){\line(0,1){10}}
\put(50.5,12){\vector(0,1){1}}\put(51,11){\makebox(1,2)[l]{$t$}}
\put(38,6){\line(1,0){9}}\put(47,6){\circle{0}}
\put(43,6){\vector(-1,0){1}}\put(45,6.5){\makebox(1,2)[l]{$(x,t)$}}
\put(43,1){\makebox(2, 1)[l]{$\gamma_{2}$}}
\end{picture}
\caption{Domain $\gamma_{1}: (x, +\infty), \gamma_{2}: (-\infty, x)$.}
\end{figure}
In addition, they are linked by the scattering matrix $s(\nu)=(s_{ij})$, namely
\begin{equation}\label{b13}
\psi_{-}(x,t,\nu)=\psi_{+}(x,t,\nu)s(\nu).
\end{equation}
\subsection{Analyticity of the Jost function }
In order to study the analytic properties of the matrix Jost
functions, we need to rewrite the spectral equation. Inserting the
representations (\ref{b11}) into the Lax pair (2.4), we get
\begin{subequations}
\begin{equation}\label{b14a}
(A^{-1}\mu_{\pm})_{x}(x,t,\nu)=\nu[\sigma,A^{-1}\mu_{\pm}(x,t,\nu)]+\Omega
 A^{-1}\mu_{\pm}(x,t,\nu),
\end{equation}
\begin{equation}\label{b14b}
(A^{-1}\mu_{\pm})_{t}(x,t,\nu)=\nu^{-1} [\sigma^{-1},A^{-1}\mu_{\pm}(x,t,\nu)]+\tilde{\Omega} A^{-1}\mu_{\pm}(x,t,\nu),
\end{equation}
\end{subequations}
where
\begin{eqnarray*}
&&\Omega(x,t,\nu)=A^{-1}\Delta{U}(x,t,\nu)A, \quad
\Delta{U}(x,t,\nu)=U(x,t,\nu)-U_{\pm}(x,t,\nu),\\
&&\tilde{\Omega}(x,t,\nu)=A^{-1}\Delta{V}(x,t,\nu)A, \quad
\Delta{V}(x,t,\nu)=V(x,t,\nu)-V_{\pm}(x,t,\nu),\\
\end{eqnarray*}
that is
\begin{eqnarray*}
&&\Omega(x,t,\nu)=\frac{1}{3}(\alpha-\alpha^{2})u_{x}\left(\begin{matrix}
0&1&-1\\
-1&0&1\\
1&-1&0
\end{matrix}\right),\\
&&\tilde{\Omega}(x,t,\nu)=\frac{1}{3\nu}\left(\begin{matrix}
2e^{u}+e^{-2u}-3&\alpha(e^{-2u}-e^{u})&\alpha^{2}(e^{-2u}-e^{u})\\
\alpha(e^{-2u}-e^{u})&\alpha^{2}(e^{-2u}+2e^{u}-3)&e^{-2u}-e^{u}\\
\alpha^{2}(e^{-2u}-e^{u})&e^{-2u}-e^{u}&\alpha(e^{-2u}+2e^{u}-3)
\end{matrix}\right).\\
\end{eqnarray*}
In the analysis that follows, we choose $t=0$ and the expressions of
$\mu_{\pm}(x,0,\nu)$ can be given from Eq. (2.14)
\begin{equation}\label{b15}
\mu_{\pm}(x,0,\nu)=A+\int_{\pm
\infty}^{x}Ae^{\nu\hat{\sigma}(x-\xi)}\Omega(\xi)A^{-1}\mu_{\pm}(\xi)d\xi,
\end{equation}
\textbf{Theorem 1.} {\sl If $u(x)\in L^{1}(-\infty,a)$ or
$L^{1}(a,+\infty)$ for any constant $a\in \mathbb{R}$, the columns
of $\mu_{\pm}(x,0,\nu)$ can be analytically extended into the
corresponding regions of the complex $\nu$-plane:
\begin{equation}\label{b16}
\begin{aligned}
 \mu_{+1}(x,0,\nu): ~\nu\in D_{3}\cup D_{4},  \quad \mu_{-1}(x,0,\nu): ~\nu\in D_{1}\cup D_{6},\\
 \mu_{+2}(x,0,\nu): ~\nu\in D_{1}\cup D_{2}, \quad \mu_{-2}(x,0,\nu): ~\nu\in D_{4}\cup D_{5},\\
 \mu_{+3}(x,0,\nu): ~\nu\in D_{5}\cup D_{6}, \quad \mu_{-3}(x,0,\nu): ~\nu\in D_{2}\cup D_{3},
\end{aligned}
\end{equation}
where the subscript $j$ refer to the columns and
$D_{n}(n=1,2,\cdots,6)$ are defined by
\begin{equation}\label{b17}
\begin{aligned}
D_{1}=\Bigg\{\nu\in\mathbb{C}|\Re l_{2}<\Re l_{3}<\Re l_{1}\Bigg\},\
D_{2}=\Bigg\{\nu\in\mathbb{C}|\Re l_{2}<\Re l_{1}<\Re l_{3}\Bigg\},\\
D_{3}=\Bigg\{\nu\in\mathbb{C}|\Re l_{1}<\Re l_{2}<\Re l_{3}\Bigg\},\
D_{4}=\Bigg\{\nu\in\mathbb{C}|\Re l_{1}<\Re l_{3}<\Re l_{2}\Bigg\},\\
D_{5}=\Bigg\{\nu\in\mathbb{C}|\Re l_{3}<\Re l_{1}<\Re l_{2}\Bigg\},\
D_{6}=\Bigg\{\nu\in\mathbb{C}|\Re l_{3}<\Re l_{2}<\Re l_{1}\Bigg\},\\
\end{aligned}
\end{equation}
namely
\begin{equation*}
D_{n}=\Bigg\{\nu\in\mathbb{C}:
\arg\nu\in\bigg(\frac{(n-1)\pi}{3},\frac{n\pi}{3}\bigg
)\Bigg\}, \quad n=1, \ldots, 6.
\end{equation*}
Here $l_{i}(\nu)(i=1,2,3)$ are the diagonal entries of matrices
$\nu\sigma$.
}\\
{\bf Proof}~~ Setting $J_{\pm}(x,0,\nu)=A^{-1}\mu_{\pm}(x,0,\nu)$,
(2.15) can be rewritten in the column form
\begin{subequations}
\begin{equation}\label{b18a}
 J_{\pm 1}(x,0,\nu)=\left(\begin{array}{c}
1\\
0\\
0
\end{array}\right)+\int_{\pm\infty}^{x}e^{\nu\sigma(x-\xi)}\Omega(\xi)e^{-\nu(x-\xi)}J_{\pm1}(\xi,t,\nu)d\xi,
\end{equation}
\begin{equation}\label{b18b}
J_{\pm 2}(x,0,\nu)=\left(\begin{array}{c}
0\\
1\\
0
\end{array}\right)+\int_{\pm
\infty}^{x}e^{\nu\sigma(x-\xi)}\Omega(\xi)e^{-\alpha\nu(x-\xi)}J_{\pm2}(\xi,t,\nu)d\xi,
\end{equation}
\begin{equation}\label{b18c}
J_{\pm 3}(x,0,\nu)=\left(\begin{array}{c}
0\\
0\\
1
\end{array}\right)+\int_{\pm
\infty}^{x}e^{\nu\sigma(x-\xi)}\Omega(\xi)e^{-\alpha^{2}\nu(x-\xi)}J_{\pm3}(\xi,t,\nu)d\xi.
\end{equation}
\end{subequations}
We consider the integral equation of $J_{+1}(x,0,\nu)$, that is
\begin{equation}\label{b19}
J_{+1}(x,0,\nu)=\left(\begin{array}{c}
1\\
0\\
0
\end{array}\right)+\int_{+
\infty}^{x}\kappa_{+}(x,\xi,\nu)J_{+1}(\xi,\nu)d\xi,
\end{equation}
where $\kappa_{+}(x,\xi,\nu)=e^{\nu(x-\xi)\sigma}e^{-\nu(x-\xi)}A^{-1}\tilde{U}A$. Then $\|\kappa_{+}(x,\xi,\nu)\|\leq 6\rho |u_{x}|, \nu\in D_{3}\cup D_{4}$, where $\rho= \max\{\|A\|, \|A\|^{-1}\}, \|\Delta{U}\|= 2|u_{x}|$.

Furthermore, we define the Neumann series as
$$J_{+1}(x,0,\nu)=\displaystyle\sum_{j=0}^{\infty}C^{(j)}(x,0,\nu)
, \quad\|C^{(j)}(x,0,\nu)\|\leq\frac{\mu^{j}(x)}{j!},$$
where
\begin{eqnarray*}
&&\mu(x)=6\rho\int_{+\infty}^{x} |u_{\xi}|d\xi,\quad
C^{(0)}(x,0,\nu)=\left(\begin{array}{c}
1\\
0\\
0
\end{array}\right), \\
&&C^{(j+1)}(x,0,\nu)=\int_{+\infty}^{x}
e^{\nu\sigma(x-\xi)}\Omega(\xi)e^{-\nu(x-\xi)}C^{(j)
}(\xi,t,\nu)d\xi.
\end{eqnarray*}
This prove that $J_{+1}(x,0,\nu)$ is analytic in domain $\nu\in
D_{3}\cup D_{4}$. The others can be proved similarly. \qquad$\square$

Under the same hypotheses as in Theorem 1, the scattering
coefficients can be analytically extended from $\Sigma$ (the contour
$\Sigma$ is defined as the boundary of domain $D$  i.e
$\Sigma=\Sigma_{1}\cup\Sigma_{2}\cup\Sigma_{3}\cup\Sigma_{4}
\cup\Sigma_{5}\cup\Sigma_{6}$ and $\Sigma_{j}=D_{j}\cap D_{j+1},
D_{6+1}=D_{1}$) to the following regions:
\begin{subequations}
\begin{equation}\label{b20a}
 s_{11}: \nu\in D_{1}\cup D_{6},\quad s_{22}: \nu\in D_{4}\cup D_{5}, \quad s_{33}: \nu\in D_{2}\cup D_{3},
\end{equation}
\begin{equation}\label{b20b}
r_{11}: \nu\in D_{3}\cup D_{4}, \quad r_{22}: \nu\in D_{1}\cup
D_{2},\quad  r_{33}: \nu\in D_{5}\cup D_{6},
\end{equation}
\end{subequations}
where $s_{ij}=(s(\nu))_{ij}$ and $r_{ij}=(s^{-1}(\nu))_{ij}$.
\begin{figure}
\setlength{\unitlength}{0.06in}
\begin{picture}(0,20)
\put(30,15){\line(1,0){30}}
\put(35,15){\vector(1,0){1}}\put(30.2,15.5){\makebox(2,1)[l]{$\pi$}}
\put(55,15){\vector(-1,0){1}}\put(59.8,15.5){\makebox(2,1)[l]{$0$}}
\put(45,0){\line(0,1){30}}\put(45,5){\vector(0,1){1}}\put(45,25){\vector(0,-1){1}}
\put(45.2,27.5){\makebox(1,2)[l]{$\frac{\pi}{2}$}}
\put(45.2,0.5){\makebox(1,2)[l]{$\frac{3\pi}{2}$}}
\put(38,1){\line(1,2){13.5}}
\put(42,9){\vector(-1,-2){1}}\put(37.2,3.5){\makebox(1,2)[l]{$\frac{4\pi}{3}$}}
\put(48,21){\vector(1,2){1}}\put(51.2,25.5){\makebox(1,2)[l]{$\frac{\pi}{3}$}}
\put(38,29){\line(1,-2){13}}
\put(42,21){\vector(-1,2){1}}\put(36,25.5){\makebox(1,2)[l]{$\frac{2\pi}{3}$}}
\put(48,9){\vector(1,-2){1}}\put(50.2,4.5){\makebox(1,2)[l]{$\frac{5\pi}{3}$}}
\end{picture}
\caption{Domain $D^+$, $D^-$ and their boundary curve $\partial D$.}
\end{figure}
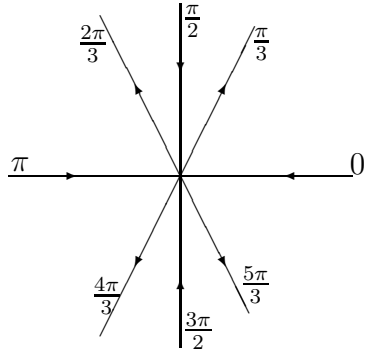
Next, we define six solutions $M_{n}(x,0,\nu)$
of (2.14), which take the following form
\begin{equation}\label{b21}
(M_{n})_{ij}(x,0,\nu)=\delta_{ij}+\int_{\gamma_{ij}^{n}}\Big(e^{\nu\hat{\sigma}
(x-x^{\prime})}\Omega(x^{\prime},0,\nu)M_{n}(x^{\prime},0,\nu)
dx^{\prime}\Big)_{ij},
\end{equation}
where the contours $\gamma_{ij}^{n}$, $(n=1, \ldots, 6, i, j=1, 2,
3)$ are defined as
\begin{equation*}
\gamma_{ij}^{n}=\left\{\begin{array}{ll}\gamma_{1},\,&\Re(l_{i})>\Re(l_{j}),
\\\gamma_{2},\,&\Re(l_{i})<\Re(l_{j}) ,
\end{array}\right.
\end{equation*}
and $\gamma_{2}$: $(-\infty,x)$, $\gamma_{1}$: $(x,+\infty). $

For concretely
\begin{equation*}
\begin{aligned}
\gamma^{1}=\left(\begin{matrix}
\gamma_{2}&\gamma_{1}&\gamma_{1}\\
\gamma_{2}&\gamma_{2}&\gamma_{2}\\
\gamma_{2}&\gamma_{1}&\gamma_{2}
\end{matrix}\right), \quad \gamma^{2}=\left(\begin{matrix}
\gamma_{2}&\gamma_{1}&\gamma_{2}\\
\gamma_{2}&\gamma_{2}&\gamma_{2}\\
\gamma_{1}&\gamma_{1}&\gamma_{2}
\end{matrix}\right), \\
\gamma^{3}=\left(\begin{matrix}
\gamma_{2}&\gamma_{2}&\gamma_{2}\\
\gamma_{1}&\gamma_{2}&\gamma_{2}\\
\gamma_{1}&\gamma_{1}&\gamma_{2}
\end{matrix}\right), \quad \gamma^{4}=\left(\begin{matrix}
\gamma_{2}&\gamma_{2}&\gamma_{2}\\
\gamma_{1}&\gamma_{2}&\gamma_{1}\\
\gamma_{1}&\gamma_{2}&\gamma_{2}
\end{matrix}\right), \\
\gamma^{5}=\left(\begin{matrix}
\gamma_{2}&\gamma_{2}&\gamma_{1}\\
\gamma_{1}&\gamma_{2}&\gamma_{1}\\
\gamma_{2}&\gamma_{2}&\gamma_{2}
\end{matrix}\right), \quad \gamma^{6}=\left(\begin{matrix}
\gamma_{2}&\gamma_{1}&\gamma_{1}\\
\gamma_{2}&\gamma_{2}&\gamma_{1}\\
\gamma_{2}&\gamma_{2}&\gamma_{2}
\end{matrix}\right). \\
\end{aligned}
\end{equation*}

We note that $\mu_\pm(x,t,\nu)$ have the same analytic properties as
$\mu_{\pm}(x,0,\nu)$, and $M_{n}(x,t,\nu)$ can be defined in the
same way as in (\ref{b21}), where
$e^{\nu\hat{\sigma}(x-x^{\prime})}$ should be replaced by
$e^{\nu\hat{\sigma}(x-x^{\prime})+\nu^{-1}\hat\sigma t}$. Supposing
that
\begin{equation}\label{b22}
\begin{aligned}
M_{n}(x,t,\nu)=\mu_{+}(x,t,\nu)e^{\nu\hat{\sigma}x
+\nu^{-1}\hat{\sigma}^{-1}t}S_{n}(\nu)
, \\
M_{n}(x,t,\nu)=\mu_{-}(x,t,\nu)e^{\nu\hat{\sigma}x+\nu^{-1}\hat{\sigma}^{-1}t}T_{n}(\nu),
\end{aligned}
\quad \nu\in D_{n},
\end{equation}
we have
\begin{equation}\label{b23}
\begin{aligned}
&S_{n}(\nu)=s(\nu)T_{n}(\nu),\\
&S_{n}(\nu)=\lim_{x\to\infty}e^{-\nu  \hat{\sigma}x-\nu^{-1}\hat{\sigma}^{-1}t}M_{n}(x,t,\nu),\\
&T_{n}(\nu)=\lim_{x\to-\infty}e^{-\nu  \hat{\sigma}x-\nu^{-1}\hat{\sigma}^{-1}t}M_{n}(x,t,\nu),\\
\end{aligned}
\end{equation}
where $S_{n}$ can be given in terms of the entries of $s(\nu)=(s_{ij}(\nu))$ and $s^{-1}(\nu)=(r_{ij}(\nu))$ as
follows
\begin{equation*}
\begin{aligned}
S_{1}(\nu)=\left(\begin{matrix}
s_{11}(\nu)&0&0\\
s_{21}(\nu)&\frac{1}{r_{22}(\nu)}&\frac{-r_{23}(\nu)}{s_{11}(\nu)}\\
s_{31}(\nu)&0&\frac{r_{22}(\nu)}{s_{11}(\nu)}
\end{matrix}\right), \quad S_{2}(\nu)=\left(\begin{matrix}
\frac{r_{22}(\nu)}{s_{33}(\nu)}&0&s_{13}(\nu)\\
\frac{-r_{21}(\nu)}{s_{33}(\nu)}&\frac{1}{r_{22}(\nu)}&s_{23}(\nu)\\
0&0&s_{33}(\nu)
\end{matrix}\right), \\
S_{3}(\nu)=\left(\begin{matrix}
\frac{1}{r_{11}(\nu)}&\frac{-r_{12}(\nu)}{s_{33}(\nu)}&s_{13}(\nu)\\
0&\frac{r_{11}(\nu)}{s_{33}(\nu)}&s_{23}(\nu)\\
0&0&s_{33}(\nu)
\end{matrix}\right), \quad S_{4}(\nu)=\left(\begin{matrix}
\frac{1}{r_{11}(\nu)}&s_{12}(\nu)&\frac{-r_{13}(\nu)}{s_{22}(\nu)}\\
0&s_{22}(\nu)&0\\
0&s_{32}(\nu)&\frac{r_{11}(\nu)}{s_{22}(\nu)}
\end{matrix}\right), \\
S_{5}(\nu)=\left(\begin{matrix}
\frac{r_{33}(\nu)}{s_{22}(\nu)}&s_{12}(\nu)&0\\
0&s_{22}(\nu)&0\\
-\frac{r_{31}(\nu)}{s_{22}(\nu)}&s_{32}(\nu)&\frac{1}{r_{33}(\nu)}
\end{matrix}\right), \quad S_{6}(\nu)=\left(\begin{matrix}
s_{11}(\nu)&0&0\\
s_{21}(\nu)&\frac{r_{33}(\nu)}{s_{11}(\nu)}&0\\
s_{31}(\nu)&\frac{-r_{32}(\nu)}{s_{11}(\nu)}&\frac{1}{r_{33}(\nu)}
\end{matrix}\right). \\
\end{aligned}
\end{equation*}

\subsection{Adjoint problem and auxiliary eigenfunctions}
It is remarked that, for the Tzitz\'{e}ica equation (\ref{a1}),
traditional inverse scattering transform may fail to construct a set
of analytic eigenfunctions in any given domain. One efficient
technique is to introduce the adjoint Lax pair
\cite{jmp47-063508,sam55-9}
\begin{subequations}
\begin{equation}\label{b24a}
\tilde{\psi}_{x}=\tilde{U}(x,\nu,u)\tilde{\psi},
\end{equation}
\begin{equation}\label{b24b}
\tilde{\psi}_{t}=\tilde{V}(x,\nu,u)\tilde{\psi},
\end{equation}
\end{subequations}
where
$$\tilde{U}(x,\nu,u)=U^{\ast}(x,-\nu^{\ast},u), \quad
\tilde{V}(x,\nu,u)=V^{\ast}(x,-\nu^{\ast},u), $$ and the star
denotes the complex conjugate.

We note that if $\phi(x,t,\nu)$ and $\varphi(x,t,\nu)$ are two arbitrary vector
eigenfunctions of the adjoint problem (2.24), then
\begin{equation}\label{b25}
\frac{T_{1}}{3\gamma^{\ast}}[\phi\times\varphi](x,t,\nu), \quad T_{1}=\left(\begin{matrix}
0&1&0\\
1&0&0\\
0&0&1
\end{matrix}\right),
\end{equation}
is an eigenfunction of the linear system (2.4). Here the symbol
$\times$ denotes the usual cross product.

Suppose $\tilde{\psi}_{\pm}(x,t,\nu)$ defined similarly as in
(\ref{b9}) are fundamental matrix solutions of the (2.24a), where
$\tilde{A}$ should be $\tilde{A}=A^{\ast}$, then there exists an
invertible $3\times3$ matrix $\tilde{s}(\nu)$ such that
\begin{equation*}
\tilde{\psi}_{-}(x,t,\nu)=\tilde{\psi}_{+}(x,t,\nu)\tilde{s}(\nu).
\end{equation*}
Introducing the modified adjoint eigenfunction $\tilde{\mu}_\pm(x,t,\nu)$ defined by
$$\tilde{\psi}_\pm=\tilde{\mu}_\pm(x,t,\nu)e^{-\nu\sigma^{-1}x-\nu^{-1}\sigma t},$$
and proceeding the same analysis for $\mu_\pm(x,t,\nu)$, we know that different collum of $\tilde{\mu}_\pm(x,t,\nu)$
can be analyticity extended into the following domains
\begin{equation}\label{b26}
\begin{aligned}
\tilde{\mu}_{+1}(x,t,\nu): ~\nu\in D_{1}\cup D_{6}, \quad \tilde{\mu}_{-1}(x,t,\nu): ~\nu\in D_{3}\cup D_{4},\\
\tilde{\mu}_{+2}(x,t,\nu): ~\nu\in D_{2}\cup D_{3}, \quad \tilde{\mu}_{-2}(x,t,\nu): ~\nu\in D_{5}\cup D_{6},\\
\tilde{\mu}_{+3}(x,t,\nu): ~\nu\in D_{4}\cup D_{5}, \quad \tilde{\mu}_{-3}(x,t,\nu): ~\nu\in D_{1}\cup D_{2}.
\end{aligned}
\end{equation}
Then the associated scattering coefficients can be similarly extended in the following region
\begin{subequations}
\begin{equation}\label{b27a}
 \tilde{s}_{11}: \nu\in D_{3}\cup D_{4}, \quad\tilde{s}_{22}: \nu\in D_{5}\cup D_{6}, \quad\tilde{s}_{33}: \nu\in D_{1}\cup D_{2},
\end{equation}
\begin{equation}\label{b27b}
\tilde{r}_{11}: \nu\in D_{1}\cup D_{6}, \quad\tilde{r}_{22}: \nu\in D_{2}\cup D_{3}, \quad\tilde{r}_{33}: \nu\in D_{4}\cup D_{5}.
\end{equation}
\end{subequations}
Using (\ref{b25}) and the solutions $\tilde{\psi}_\pm$ of the adjoint problem (2.24), we construct six novel solutions for
the Lax pair (2.4) as
\begin{subequations}
\begin{equation}\label{b28a}
\chi_{1}(x,t,\nu)=\frac{T_{1}}{3\gamma^{\ast}}[\tilde{\psi}_{-3}\times\tilde{\psi}_{+1}](x,t,\nu),
\end{equation}
\begin{equation}\label{b28b}
\chi_{2}(x,t,\nu)=\frac{T_{1}}{3\gamma^{\ast}}[\tilde{\psi}_{+2}\times\tilde{\psi}_{-3}](x,t,\nu),
\end{equation}
\begin{equation}\label{b28c}
\chi_{3}(x,t,\nu)=\frac{T_{1}}{3\gamma^{\ast}}[\tilde{\psi}_{-1}\times\tilde{\psi}_{+2}](x,t,\nu),
\end{equation}
\begin{equation}\label{b28d}
\chi_{4}(x,t,\nu)=\frac{T_{1}}{3\gamma^{\ast}}[\tilde{\psi}_{+3}\times\tilde{\psi}_{-1}](x,t,\nu),
\end{equation}
\begin{equation}\label{b28e}
\chi_{5}(x,t,\nu)=\frac{T_{1}}{3\gamma^{\ast}}[\tilde{\psi}_{-2}\times\tilde{\psi}_{+3}](x,t,\nu),
\end{equation}
\begin{equation}\label{b28f}
\chi_{6}(x,t,\nu)=\frac{T_{1}}{3\gamma^{\ast}}
[\tilde{\psi}_{+1}\times\tilde{\psi}_{-2}](x,t,\nu).
\end{equation}
\end{subequations}
 Here, $\chi_{1}(x,t,\nu), \cdots, \chi_{6}(x,t,\nu)$ are called the auxiliary eigenfunctions.
Due to the Tzitz\'{e}ica equation has six different domains of
analyticity, we need to define six auxiliary eigenfunctions.
Furthermore, $\chi_{n}(x,t,\nu)$ is analytic in domain $\nu\in
D_{n}, n=1, \cdots, 6$.

In addition, for $\nu\in \Sigma$, the Jost functions $\psi_\pm$ in (\ref{b9}) and the solutions $\tilde{\psi}_\pm$ of the adjoint problem (2.24) satisfy the following relations
\begin{subequations}
\begin{equation}\label{b29a}
\psi_{\pm1}(\nu)=\frac{T_{1}}{3\gamma^{\ast}}[\tilde{\psi}_{\pm2}\times\tilde{\psi}_{\pm3}](x,t,\nu),
 \end{equation}
\begin{equation}\label{b29b}
\psi_{\pm2}(\nu)=\frac{T_{1}}{3\gamma^{\ast}}[\tilde{\psi}_{\pm1}\times\tilde{\psi}_{\pm2}](x,t,\nu),
 \end{equation}
\begin{equation}\label{b29c}
\psi_{\pm3}(\nu)=\frac{T_{1}}{3\gamma^{\ast}}[\tilde{\psi}_{\pm3}\times\tilde{\psi}_{\pm1}](x,t,\nu),
\end{equation}
\begin{equation}\label{b29d}
\tilde{\psi}_{\pm1}(\nu)=\frac{T_{1}}{3\gamma}[\psi_{\pm2}\times \psi_{\pm3}](x,t,\nu),
\end{equation}
\begin{equation}\label{b29e}
\tilde{\psi}_{\pm2}(\nu)=\frac{T_{1}}{3\gamma}[\psi_{\pm1}\times \psi_{\pm2}](x,t,\nu),
\end{equation}
\begin{equation}\label{b29f}
\tilde{\psi}_{\pm3}(\nu)=\frac{T_{1}}{3\gamma}[\psi_{\pm3}\times \psi_{\pm1}](x,t,\nu).
\end{equation}
\end{subequations}
Based on the above facts, the scattering matrices $s(\nu)$ and $\tilde{s}(\nu)$ are related by
\begin{equation}\label{b30}
\tilde{s}(\nu)=\Gamma\bigg(s^{-1}(\nu)\bigg)^{T}\Gamma, \quad \Gamma=\left(\begin{matrix}
1&0&0\\
0&0&1\\
0&1&0
\end{matrix}\right).
\end{equation}
For all $\nu\in\Sigma$, the Jost functions and the auxiliary eigenfunctions obey the following equations
\begin{subequations}
\begin{equation}\label{b31a}
\psi_{+1}(\nu)=\frac{1}{r_{22}(\nu)}[\chi_{2}(\nu)+
\psi_{+2}(\nu)r_{21}(\nu)]=\frac{1}{r_{33}(\nu)}[\chi_{5}(\nu)+
\psi_{+3}(\nu)r_{31}(\nu)],
\end{equation}
\begin{equation}\label{b31b}
\psi_{+2}(\nu)=\frac{1}{r_{11}(\nu)}[\chi_{3}(\nu)+
\psi_{+1}(\nu)r_{12}(\nu)]=\frac{1}{r_{33}(\nu)}[\chi_{6}(\nu)+
\psi_{+3}(\nu)r_{32}(\nu)],
\end{equation}
\begin{equation}\label{b31c}
\psi_{+3}(\nu)=\frac{1}{r_{22}(\nu)}[\chi_{1}(\nu)+
\psi_{+2}(\nu)r_{23}(\nu)]=\frac{1}{r_{11}(\nu)}[\chi_{4}(\nu)+
\psi_{+1}(\nu)r_{13}(\nu)],
\end{equation}
\begin{equation}\label{b31d}
\psi_{-1}(\nu)=\frac{1}{s_{33}(\nu)}[\chi_{2}(\nu)+
\psi_{-3}(\nu)s_{31}(\nu)]=\frac{1}{s_{22}(\nu)}[\chi_{5}(\nu)+
\psi_{-2}(\nu)s_{21}(\nu)],
\end{equation}
\begin{equation}\label{b31e}
\psi_{-2}(\nu)=\frac{1}{s_{33}(\nu)}[\chi_{3}(\nu)+
\psi_{-3}(\nu)s_{32}(\nu)]=\frac{1}{s_{11}(\nu)}[\chi_{6}(\nu)+
\psi_{-1}(\nu)s_{12}(\nu)],
\end{equation}
\begin{equation}\label{b31f}
\psi_{-3}(\nu)=\frac{1}{s_{11}(\nu)}[\chi_{1}(\nu)+
\psi_{-1}(\nu)s_{13}(\nu)]=\frac{1}{s_{22}(\nu)}[\chi_{4}(\nu)+
\psi_{-2}(\nu)s_{23}(\nu)].
\end{equation}
\end{subequations}
\subsection{symmetries}
\textit{First symmetry.} It is readily verified that if $\psi(x,t,\nu)$ is a fundamental matrix
solution of the Lax pair (2.4), so is
$T_{1}\bigg(\psi^{\dagger}(x,t,-\nu^{\ast})\bigg)^{-1}$, where
$T_{1}$ is defined in (2.25).

For all $\nu\in\Sigma$, the Jost functions satisfy the symmetry
\begin{equation*}
T_{1}\bigg(\psi_{\pm}^{\dagger}(x,t,-\nu^{\ast})\bigg)^{-1}\frac{\Gamma}{3}=\psi_{\pm}(x,t,\nu),
\end{equation*}
where
$$\bigg(\psi^{-1}_{\pm}(x,t,\nu)\bigg)^{T}=
\frac{1}{\det \psi_{\pm}(x,t,\nu)}\bigg(\psi_{\pm2}\times
\psi_{\pm3}, \psi_{\pm3}\times \psi_{\pm1}, \psi_{\pm1}\times
\psi_{\pm2}\bigg)(x,t,\nu). $$
This symmetry implies a condition between the scattering matrix and its inverse
\begin{equation}\label{b32}
s^{-1}(\nu)=\Gamma s^{\dagger}(-\nu^{\ast})\Gamma^{-1}, \quad
\nu\in\Sigma,
\end{equation}
or, in component form
\begin{subequations}
\begin{equation}\label{b33a}
r_{11}(\nu)=s^{\ast}_{11}(-\nu^{\ast}), \quad r_{12}(\nu)=s^{\ast}_{31}(-\nu^{\ast}), \quad r_{13}(\nu)=s^{\ast}_{21}(-\nu^{\ast}),
\end{equation}
\begin{equation}\label{b33b}
r_{21}(\nu)=s^{\ast}_{13}(-\nu^{\ast}), \quad r_{22}(\nu)=s^{\ast}_{33}(-\nu^{\ast}), \quad
r_{23}(\nu)=s^{\ast}_{23}(-\nu^{\ast}),
\end{equation}
\begin{equation}\label{b33c}
r_{31}(\nu)=s^{\ast}_{12}(-\nu^{\ast}), \quad r_{32}(\nu)=s^{\ast}_{32}(-\nu^{\ast}), \quad
r_{33}(\nu)=s^{\ast}_{22}(-\nu^{\ast}).
\end{equation}
\end{subequations}
The Schwarz reflection principle of (\ref{b32}) gives to the following results
\begin{subequations}
\begin{equation}\label{b34a}
r_{11}(\nu)=s^{\ast}_{11}(-\nu^{\ast}), \quad \nu\in D_{3}\cup D_{4},
\end{equation}
\begin{equation}\label{b34b}
r_{22}(\nu)=s^{\ast}_{33}(-\nu^{\ast}), \quad \nu\in D_{1}\cup D_{2},
\end{equation}
\begin{equation}\label{b34c}
r_{33}(\nu)=s^{\ast}_{22}(-\nu^{\ast}), \quad \nu\in D_{5}\cup D_{6}.
\end{equation}
\end{subequations}

The Jost functions and the auxiliary eigenfunctions obey the following symmetry relations
\begin{subequations}
\begin{equation}\label{b35a}
\psi^{\ast}_{+1}(-\nu^{\ast})=\frac{T_{1}}{3\gamma r_{33}(\nu)}[\chi_{6}(\nu)\times
\psi_{+3}(\nu)]=\frac{T_{1}}{3\gamma r_{22}(\nu)}[
\psi_{+2}(\nu)\times\chi_{1}(\nu)],
\end{equation}
\begin{equation}\label{b35b}
\psi^{\ast}_{-1}(-\nu^{\ast})=\frac{T_{1}}{3\gamma s_{33}(\nu)}[\chi_{3}(\nu)\times
\psi_{-3}(\nu)]=\frac{T_{1}}{3\gamma s_{22}(\nu)}[
\psi_{-2}(\nu)\times\chi_{4}(\nu)],
 \end{equation}
\begin{equation}\label{b35c}
\psi^{\ast}_{+2}(-\nu^{\ast})=\frac{T_{1}}{3\gamma r_{22}(\nu)}[\chi_{2}(\nu)\times
\psi_{+2}(\nu)]=\frac{T_{1}}{3\gamma r_{11}(\nu)}[
\psi_{+1}(\nu)\times\chi_{3}(\nu)],
\end{equation}
\begin{equation}\label{b35d}
\psi^{\ast}_{-2}(-\nu^{\ast})=\frac{T_{1}}{3\gamma s_{22}(\nu)}[\chi_{5}(\nu)\times
\psi_{-2}(\nu)]=\frac{T_{1}}{3\gamma s_{11}(\nu)}[
\psi_{-1}(\nu)\times\chi_{6}(\nu)],
\end{equation}
\begin{equation}\label{b35e}
\psi^{\ast}_{+3}(-\nu^{\ast})=\frac{T_{1}}{3\gamma r_{11}(\nu)}[\chi_{4}(\nu)\times
\psi_{+1}(\nu)]=\frac{T_{1}}{3\gamma r_{33}(\nu)}[
\psi_{+3}(\nu)\times\chi_{5}(\nu)],
\end{equation}
\begin{equation}\label{b35f}
\psi^{\ast}_{-3}(-\nu^{\ast})=\frac{T_{1}}{3\gamma
s_{11}(\nu)}[\chi_{1}(\nu)\times
\psi_{-1}(\nu)]=\frac{T_{1}}{3\gamma
s_{33}(\nu)}[\psi_{-3}(\nu)\times\chi_{2}(\nu)].
\end{equation}
\end{subequations}

Besides, the auxiliary eigenfunctions and the Jost functions satisfy another symmetry conditions
\begin{subequations}
\begin{equation}\label{b36a}
\chi^{\ast}_{1}(x,t,-\nu^{\ast})=\frac{T_{1}}{3\gamma}[\psi_{-3}\times
\psi_{+1}](x,t,\nu), \quad \nu\in D_{3},
\end{equation}
\begin{equation}\label{b36b}
\chi^{\ast}_{2}(x,t,-\nu^{\ast})=\frac{T_{1}}{3\gamma}[\psi_{+2}\times
\psi_{-3}](x,t,\nu), \quad\nu\in D_{2},
 \end{equation}
\begin{equation}\label{b36c}
\chi^{\ast}_{3}(x,t,-\nu^{\ast})=\frac{T_{1}}{3\gamma}[\psi_{-1}\times
\psi_{+2}](x,t,\nu), \quad \nu\in D_{1},
\end{equation}
\begin{equation}\label{b36d}
\chi^{\ast}_{4}(x,t,-\nu^{\ast})=\frac{T_{1}}{3\gamma}[\psi_{+3}\times
\psi_{-1}](x,t,\nu), \quad \nu\in D_{6},
\end{equation}
\begin{equation}\label{b36e}
\chi^{\ast}_{5}(x,t,-\nu^{\ast})=\frac{T_{1}}{3\gamma}[\psi_{-2}\times
\psi_{+3}](x,t,\nu), \quad \nu\in D_{5},
\end{equation}
\begin{equation}\label{b36f}
\chi^{\ast}_{6}(x,t,-\nu^{\ast})=\frac{T_{1}}{3\gamma}[\psi_{+1}\times
\psi_{-2}](x,t,\nu), \quad \nu\in D_{4}.
\end{equation}
\end{subequations}
\textit{Second symmetry.} We note that if $\psi_{\pm}(x,t,\nu)$ is a solution of
the Lax pair (2.4), so is $\psi_{\pm}(x,t,\alpha^{2}\nu)$.

For all $\nu\in\Sigma$, the Jost eigenfunctions satisfy the symmetry
\begin{equation}\label{b37}
\psi_{\pm}(\nu)=\alpha\sigma\psi_{\pm}(\alpha^{2}\nu)\Delta,\quad \Delta=\left(\begin{matrix}
0&0&1\\
1&0&0\\
0&1&0
\end{matrix}\right).
\end{equation}
In component form, (\ref{b37}) yields
\begin{subequations}
\begin{equation}\label{b38a}
\psi_{\pm1}(\nu)=\alpha\sigma\psi_{\pm2}(\alpha^{2}\nu), \quad \nu\in \Re l_{1}\lessgtr\Re l_{j}, \quad j=2, 3,
\end{equation}
\begin{equation}\label{b38b}
\psi_{\pm2}(\nu)=\alpha\sigma\psi_{\pm3}(\alpha^{2}\nu), \quad \nu\in \Re l_{2}\lessgtr\Re l_{j}, \quad j=1, 3,
\end{equation}
\begin{equation}\label{b38c}
\psi_{\pm3}(\nu)=\alpha\sigma\psi_{\pm1}(\alpha^{2}\nu), \quad \nu\in \Re l_{3}\lessgtr\Re l_{j}, \quad j=1, 2.
\end{equation}
\end{subequations}
The scattering matrix satisfies the symmetry
\begin{equation}\label{b39}
s(\alpha^{2}\nu)=\Delta s(\nu)\Delta^{-1}, \quad \nu\in\Sigma.
\end{equation}
Componentwise, we have
\begin{subequations}
\begin{equation}\label{b40a}
s_{11}(\nu)=s_{22}(\alpha^{2}\nu), \quad
s_{12}(\nu)=s_{23}(\alpha^{2}\nu), \quad
s_{13}(\nu)=s_{21}(\alpha^{2}\nu),
\end{equation}
\begin{equation}\label{b40b}
s_{21}(\nu)=s_{32}(\alpha^{2}\nu), \quad
s_{22}(\nu)=s_{33}(\alpha^{2}\nu), \quad
s_{23}(\nu)=s_{31}(\alpha^{2}\nu),
 \end{equation}
\begin{equation}\label{b40c}
s_{31}(\nu)=s_{12}(\alpha^{2}\nu), \quad
s_{32}(\nu)=s_{13}(\alpha^{2}\nu), \quad
s_{33}(\nu)=s_{11}(\alpha^{2}\nu).
\end{equation}
\end{subequations}
Beside, the analyticity properties of the scattering matrix imply
that
\begin{subequations}
\begin{equation}\label{b41a}
\begin{aligned}
s_{11}(\nu)=s_{22}(\alpha^{2}\nu), ~\nu\in D_{1}\cup D_{6}, \quad
s_{22}(\nu)=s_{33}(\alpha^{2}\nu), ~\nu\in D_{4}\cup D_{5}, \\
s_{33}(\nu)=s_{11}(\alpha^{2}\nu), ~\nu\in D_{2}\cup D_{3},
\end{aligned}
\end{equation}
\begin{equation}\label{b41b}
\begin{aligned}
r_{11}(\nu)=r_{22}(\alpha^{2}\nu), ~\nu\in D_{3}\cup D_{4}, \quad
r_{22}(\nu)=r_{33}(\alpha^{2}\nu), ~\nu\in D_{1}\cup D_{2},\\
r_{33}(\nu)=r_{11}(\alpha^{2}\nu), ~\nu\in D_{5}\cup D_{6}.
\end{aligned}
\end{equation}
\end{subequations}
Finally, taking (2.31), (2.37) and (2.39) into consideration, we find
\begin{subequations}
\begin{equation}\label{b42a}
\chi_{1}(\nu)=\alpha\sigma\chi_{5}(\alpha^{2}\nu)=\alpha^{2}\sigma^{2}\chi_{3}(\alpha\nu), \quad \nu\in D_{1},
\end{equation}
\begin{equation}\label{b42b}
\chi_{2}(\nu)=\alpha\sigma\chi_{6}(\alpha^{2}\nu)=\alpha^{2}\sigma^{2}\chi_{4}(\alpha\nu), \quad
\nu\in D_{2}.
\end{equation}
\end{subequations}
\textit{Third symmetry.} 
It is the fact that $\psi_{\pm}(x,t,\nu)$ and $\psi^{\ast}_{\pm}(x,t,\nu^{\ast})$ are both the solutions of the system (2.4).

 For all $\nu\in\Sigma$, the Jost eigenfunctions satisfy the symmetry
\begin{equation}\label{b43}
\psi_{\pm}(x,t,\nu)=\psi^{\ast}_{\pm}(x,t,\nu^{\ast})\Gamma,
\end{equation}
or equivalently
\begin{subequations}
\begin{equation}\label{b44a}
\psi_{\pm1}(x,t,\nu)=\psi^{\ast}_{\pm1}(x,t,\nu^{\ast}), \quad \nu\in \Re l_{1}\lessgtr\Re l_{j}, ~j=2, 3,
 \end{equation}
\begin{equation}\label{b44b}
\psi_{\pm2}(x,t,\nu)=\psi^{\ast}_{\pm3}(x,t,\nu^{\ast}), \quad \nu\in \Re l_{2}\lessgtr\Re l_{j}, ~j=1, 3,
 \end{equation}
\begin{equation}\label{b44c}
\psi_{\pm3}(x,t,\nu)=\psi^{\ast}_{\pm2}(x,t,\nu^{\ast}), \quad \nu\in \Re l_{3}\lessgtr\Re l_{j}, ~j=1, 2.
\end{equation}
\end{subequations}
In the same way, the scattering matrices admit the symmetry
\begin{equation}\label{b45}
s^{\ast}(\nu^{\ast})=\Gamma s(\nu)\Gamma.
\end{equation}
Componentwise, we have
\begin{subequations}
\begin{equation}\label{b46a}
s^{\ast}_{11}(\nu^{\ast})=s_{11}(\nu), \quad
s^{\ast}_{12}(\nu^{\ast})=s_{13}(\nu), \quad
s^{\ast}_{13}(\nu^{\ast})=s_{12}(\nu),
\end{equation}
\begin{equation}\label{b46b}
s^{\ast}_{21}(\nu^{\ast})=s_{31}(\nu), \quad
s^{\ast}_{22}(\nu^{\ast})=s_{33}(\nu), \quad
s^{\ast}_{23}(\nu^{\ast})=s_{32}(\nu),
\end{equation}
\begin{equation}\label{b46c}
s^{\ast}_{31}(\nu^{\ast})=s_{21}(\nu), \quad
s^{\ast}_{32}(\nu^{\ast})=s_{23}(\nu), \quad
s^{\ast}_{33}(\nu^{\ast})=s_{22}(\nu).
\end{equation}
\end{subequations}
In addition, the analyticity properties of the scattering matrix give the following results
\begin{subequations}
\begin{equation}\label{b47a}
\begin{aligned}
s_{11}(\nu)=s^{\ast}_{11}(\nu^{\ast}), ~\nu\in D_{1}\cup D_{6}, \quad
s_{22}(\nu)=s^{\ast}_{33}(\nu^{\ast}), ~\nu\in D_{4}\cup D_{5}, \\
s_{33}(\nu)=s^{\ast}_{22}(\nu^{\ast}), ~\nu\in D_{2}\cup D_{3},
\end{aligned}
\end{equation}
\begin{equation}\label{b47b}
\begin{aligned}
r_{11}(\nu)=r^{\ast}_{11}(\nu^{\ast}), ~\nu\in D_{3}\cup D_{4}, \quad
r_{22}(\nu)=r^{\ast}_{33}(\nu^{\ast}), ~\nu\in D_{1}\cup D_{2},\\
r_{33}(\nu)=r^{\ast}_{22}(\nu^{\ast}), ~\nu\in D_{5}\cup D_{6}.
\end{aligned}
\end{equation}
\end{subequations}
The auxiliary eigenfunctions satisfy the symmetries
\begin{subequations}
\begin{equation}\label{b48a}
\chi_{1}(\nu)=-\frac{\gamma}{\gamma^{\ast}}\chi^{\ast}_{6}(\nu^{\ast}), \quad \nu\in D_{1},
\end{equation}
\begin{equation}\label{b48b}
\chi_{2}(\nu)=-\frac{\gamma}{\gamma^{\ast}}\chi^{\ast}_{5}(\nu^{\ast}), \quad \nu\in D_{2},
\end{equation}
\begin{equation}\label{b48c}
\chi_{3}(\nu)=-\frac{\gamma}{\gamma^{\ast}}\chi^{\ast}_{4}(\nu^{\ast}), \quad \nu\in D_{3}.
\end{equation}
\end{subequations}
The matrix spectral function $s(\nu)$ has the following properties: \\
(i) $s(\nu)=I+o(\frac{1}{\nu}), ~ s^{-1}(\nu)=I+o(\frac{1}{\nu}),
\quad \nu\longrightarrow \infty,$\\
(ii) $\det s(\nu)=1, ~\det s^{-1}(\nu)=1. $
\setcounter{equation}{0}
\section{Discrete spectrum  }
\subsection{Discrete spectrum}
In the following, we switch our attention to discuss the discrete spectrum
for the Tzitz\'{e}ica equation. To this end, we introduce the following six $3\times3$ matrices
\begin{subequations}
\begin{equation}\label{c1a}
\mathbf{G}_{1}(x,t,\nu)=\bigg(\psi_{-1}(x,t,\nu),\psi_{+2}(x,t,\nu),\chi_{1}(x,t,\nu)\bigg),
\quad \nu\in D_{1},
\end{equation}
\begin{equation}\label{c1b}
\mathbf{G}_{2}(x,t,\nu)=\bigg(\chi_{2}(x,t,\nu),\psi_{+2}(x,t,\nu),\psi_{-3}(x,t,\nu)\bigg),
\quad \nu\in D_{2},
\end{equation}
\begin{equation}\label{c1c}
\mathbf{G}_{3}(x,t,\nu)=\bigg(\psi_{+1}(x,t,\nu),\chi_{3}(x,t,\nu),\psi_{-3}(x,t,\nu)\bigg),
\quad \nu\in D_{3},
\end{equation}
\begin{equation}\label{c1d}
\mathbf{G}_{4}(x,t,\nu)=\bigg(\psi_{+1}(x,t,\nu),\psi_{-2}(x,t,\nu),\chi_{4}(x,t,\nu)\bigg),
\quad \nu\in D_{4},
\end{equation}
\begin{equation}\label{c1e}
\mathbf{G}_{5}(x,t,\nu)=\bigg(\chi_{5}(x,t,\nu),\psi_{-2}(x,t,\nu),\psi_{+3}(x,t,\nu)\bigg),
\quad \nu\in D_{5},
\end{equation}
\begin{equation}\label{c1f}
\mathbf{G}_{6}(x,t,\nu)=\bigg(\psi_{-1}(x,t,\nu),\chi_{6}(x,t,\nu),\psi_{+3}(x,t,\nu)\bigg),
\quad \nu\in D_{6}.
\end{equation}
\end{subequations}
We note that the above matrix function are analytic in their respective domain.
From (2.31), we know that
\begin{subequations}
\begin{equation}\label{c2a}
\det\Big(\mathbf{G}_{1}(x,t,\nu)\Big)=s_{11}(\nu)r_{22}(\nu)\gamma,
\quad \nu\in D_{1},
\end{equation}
\begin{equation}\label{c2b}
\det\Big(\mathbf{G}_{2}(x,t,\nu)\Big)=s_{33}(\nu)r_{22}(\nu)\gamma,\quad
\nu\in D_{2},
\end{equation}
\begin{equation}\label{c2c}
\det\Big(\mathbf{G}_{3}(x,t,\nu)\Big)=r_{11}(\nu)s_{33}(\nu)\gamma,\quad
\nu\in D_{3},
\end{equation}
\begin{equation}\label{c2d}
\det\Big(\mathbf{G}_{4}(x,t,\nu)\Big)=r_{11}(\nu)s_{22}(\nu)\gamma,\quad
\nu\in D_{4},
\end{equation}
\begin{equation}\label{c2e}
\det\Big(\mathbf{G}_{5}(x,t,\nu)\Big)=r_{33}(\nu)s_{22}(\nu)\gamma,\quad
\nu\in D_{5},
\end{equation}
\begin{equation}\label{c2f}
\det\Big(\mathbf{G}_{6}(x,t,\nu)\Big)=s_{11}(\nu)s_{33}(\nu)\gamma,\quad
\nu\in D_{6}.
\end{equation}
\end{subequations}
Thus, we can easily observe that the columns of
$\mathbf{G}_{1}(x,t,\nu)$ become linearly dependent at the zeros of
$s_{11}(\nu)$ and $r_{22}(\nu)$.

According to the symmetry conditions about the scattering matrices discussed in section 2.4, we can locate the distribution of the discrete spectrum. For convenience, we let $\nu_{0}\in D_{2}$ be a discrete eigenvalue of the scattering problem. It is easy to see that $\nu_{0}$ can be divided into two types of eigenvalue.\\
1. The first type is that
$ \nu_{0}=-\nu_{0}^{\ast}$. \\
2. The second type is that $\nu_{0}\neq-\nu_{0}^{\ast}$.

When $\nu_{0}=-\nu_{0}^{\ast}$, the symmetry conditions about the scattering matrices imply that
\begin{eqnarray}\label{c3}
&&\nonumber s_{33}(\nu_{0})=0\Leftrightarrow
s_{22}(\nu_{0}^{\ast})=0\Leftrightarrow
s_{11}(\alpha\nu_{0}^{\ast})=0\\
&&\nonumber\Leftrightarrow
s_{11}(\alpha^{2}\nu_{0})=0\Leftrightarrow
s_{22}(\alpha\nu_{0})=0  \Leftrightarrow
s_{33}(\alpha^{2}\nu_{0}^{\ast})=0\\
&&\nonumber\Leftrightarrow
r_{11}(\alpha^{2}\nu_{0}^{\ast})=0 \Leftrightarrow
r_{11}(\alpha\nu_{0})=0\Leftrightarrow r_{22}(\nu_{0})=0\\
&&\Leftrightarrow r_{22}(\alpha\nu_{0}^{\ast})=0 \Leftrightarrow
r_{33}(\nu_{0}^{\ast})=0\Leftrightarrow r_{33}(\alpha^{2}\nu_{0})=0.
\end{eqnarray}
When $\nu_{0}\neq-\nu_{0}^{\ast}$, we can derive
\begin{eqnarray}\label{c4}
&&\nonumber s_{33}(\nu_{0})=0\Leftrightarrow s_{22}(\nu_{0}^{\ast})=0\Leftrightarrow s_{11}(\alpha\nu_{0}^{\ast})=0\\
&&\nonumber\Leftrightarrow s_{11}(\alpha^{2}\nu_{0})=0 \Leftrightarrow
s_{22}(\alpha\nu_{0})=0 \Leftrightarrow s_{33}(\alpha^{2}\nu_{0}^{\ast})=0\\
&&\nonumber\Leftrightarrow r_{11}(-\alpha^{2}\nu_{0})=0
\Leftrightarrow r_{11}(-\alpha\nu_{0}^{\ast})=0\Leftrightarrow r_{22}(-\nu_{0}^{\ast})=0\\
&&\Leftrightarrow r_{22}(-\alpha\nu_{0})=0\Leftrightarrow r_{33}(-\nu_{0})=0\Leftrightarrow r_{33}(-\alpha^{2}\nu_{0}^{\ast})=0.
\end{eqnarray}

In addition, for $\nu_{0}=-\nu_{0}^{\ast}, \nu_{0}\in D_{2}$, Jost functions and the auxiliary eigenfunctions
admit the following statements\\
(1): $\chi_{2}(\nu_{0})=\textbf{0}, $\\
(2): $\chi_{4}(\alpha\nu_{0})=\textbf{0}, $\\
(3): $\chi_{6}(\alpha^{2}\nu_{0})=\textbf{0}, $\\
(4): There exists a constant $b_{2}$ such that $\psi_{+2}(-\nu_{0}^{\ast})=b_{2}\psi_{-3}(-\nu_{0}^{\ast}), $\\
(5): There exists a constant $b_{4}$ such that $\psi_{+3}(-\alpha^{2}\nu_{0}^{\ast})=b_{4}\psi_{-1}(-\alpha^{2}\nu_{0}^{\ast}), $\\
(6): There exists a constant $b_{6}$ such that $\psi_{+1}(-\alpha\nu_{0}^{\ast})=b_{6}\psi_{-2}(-\alpha\nu_{0}^{\ast}). $

Similarly, when $\nu_{0}\neq-\nu_{0}^{\ast}, \nu_{0}\in D_{2}$, the following statements are equivalent\\
(1): $\chi_{1}(\alpha\nu_{0}^{\ast})=\textbf{0}, $\\
(2): $\chi_{3}(\alpha^{2}\nu_{0}^{\ast})=\textbf{0}, $\\
(3): $\chi_{5}(\nu_{0}^{\ast})=\textbf{0}, $\\
(4): There exists a constant $b_{1}$ such that $\psi_{+1}(-\alpha^{2}\nu_{0})=b_{1}\psi_{-3}(-\alpha^{2}\nu_{0}), $\\
(5): There exists a constant $b_{3}$ such that $\psi_{+2}(-\alpha\nu_{0})=b_{3}\psi_{-1}(-\alpha\nu_{0}), $\\
(6): There exists a constant $b_{5}$ such that $\psi_{+3}(-\nu_{0})=b_{5}\psi_{-2}(-\nu_{j})$.

In what follows, we assume that the zeros of $s_{jj}(\nu)$ and
$r_{jj}(\nu)(j=1,2,3)$ are simple.

 (i) If $\nu_{0}$ is an eigenvalue of the first type, we can
directly get
$\chi_{1}(\alpha\nu_{0}^{\ast})=\chi_{2}(\nu_{0})=\textbf{0}$.  As a
result, we observe that
\begin{subequations}
\begin{equation}\label{c5a}
\psi_{+2}(\nu_{0})=f_{0}\psi_{-3}(\nu_{0}), \quad
\psi_{+1}(\alpha^{2}\nu_{0}^{\ast})=\hat{f}_{0}\psi_{-3}(\alpha^{2}\nu_{0}^{\ast}),
\end{equation}
\begin{equation}\label{c5b}
\psi_{+2}(\alpha\nu_{0}^{\ast})=\check{f}_{0}\psi_{-1}(\alpha\nu_{0}^{\ast}),
\quad
\psi_{+3}(\alpha^{2}\nu_{0})=\breve{f}_{0}\psi_{-1}(\alpha^{2}\nu_{0}),
\end{equation}
\begin{equation}\label{c5c}
\psi_{+3}(\nu_{0}^{\ast})=\acute{f}_{0}\psi_{-2}(\nu_{0}^{\ast}),
\quad
\psi_{+1}(\alpha\nu_{0})=\grave{f}_{0}\psi_{-2}(\alpha\nu_{0}),
\end{equation}
\end{subequations}
where $f_{0}, \hat{f}_{0}, \check{f}_{0}, \breve{f}_{0},
\acute{f}_{0}$, and $\grave{f}_{0}$ are the associated
proportionality constants.

 (ii) Suppose $\nu_{0}$ is an
eigenvalue of the second type. In this case, we find that
\begin{subequations}
\begin{equation}\label{c6a}
\chi_{2}(\nu_{0})=
\hat{c}_{0}\psi_{-3}(\nu_{0}), \quad \psi_{+2}(-\nu_{0}^{\ast})=\hat{d}_{0}\chi_{2}(-\nu_{0}^{\ast}),
\end{equation}
\begin{equation}\label{c6b}
\chi_{1}(\alpha\nu_{0}^{\ast})=
c_{0}\psi_{-1}(\alpha\nu_{0}^{\ast}), \quad \psi_{+2}(-\alpha\nu_{0})=d_{0}\chi_{1}(-\alpha\nu_{0}),
\end{equation}
\begin{equation}\label{c6c}
\chi_{3}(\alpha^{2}\nu_{0}^{\ast})=
\check{c}_{0}\psi_{-3}(\alpha^{2}\nu_{0}^{\ast}), \quad \psi_{+1}(-\alpha^{2}\nu_{0})=\check{d}_{0}\chi_{3}(-\alpha^{2}\nu_{0}),
\end{equation}
\begin{equation}\label{c6d}
\chi_{4}(\alpha\nu_{0})=
\breve{c}_{0}\psi_{-2}(\alpha\nu_{0}), \quad \psi_{+1}(-\alpha\nu_{0}^{\ast})=\breve{d}_{0}\chi_{4}(-\alpha\nu_{0}^{\ast}),
\end{equation}
\begin{equation}\label{c6e}
\chi_{5}(\nu_{0}^{\ast})=
\acute{c}_{0}\psi_{-2}(\nu_{0}^{\ast}), \quad \psi_{+3}(-\nu_{0})=\acute{d}_{0}\chi_{5}(-\nu_{0}),
\end{equation}
\begin{equation}\label{c6f}
\chi_{6}(\alpha^{2}\nu_{0})=
\grave{c}_{0}\psi_{-1}(\alpha^{2}\nu_{0}), \quad
\psi_{+3}(-\alpha^{2}\nu_{0}^{\ast})=\grave{d}_{0}\chi_{6}(-\alpha^{2}\nu_{0}^{\ast}),
\end{equation}
\end{subequations}
where, $d_{0}, \hat{d}_{0}, \check{d}_{0}, \breve{d}_{0},
\acute{d}_{0}$, $\grave{d}_{0}$, $c_{0}, \hat{c}_{0}, \check{c}_{0},
\breve{c}_{0}, \acute{c}_{0}$ and $\grave{c}_{0}$ are  the
associated proportionality constants.
\subsection{Symmetries of the norming constants}
To obtain the residue conditions for the inverse problem, we rewrite
the equations (3.5) and (3.6) in terms of the modified
eigenfunctions. Setting $\{\zeta_{j}\}_{j=1}^{N_{1}}$ be the set of
all eigenvalues of the first type, then from (\ref{b11}) and (3.5),
we have
\begin{subequations}
\begin{equation}\label{c7a}
\mu_{+2}(\zeta_{j})=f_{j}\mu_{-3}(\zeta_{j})e^{(\alpha^{2}-\alpha)
\zeta_{j}x+(\alpha-\alpha^{2})\ddot{\zeta_{j}}t},
\end{equation}
\begin{equation}\label{c7b}
\mu_{+1}(\alpha^{2}\zeta_{j}^{\ast})=\hat{f}_{j}
\mu_{-3}(\alpha^{2}\zeta_{j}^{\ast})
e^{(\alpha-\alpha^{2})\zeta_{j}^{\ast}x+(\alpha^{2}-\alpha)\ddot{\zeta_{j}}^{\ast}t},
\end{equation}
\begin{equation}\label{c7c}
\mu_{+2}(\alpha\zeta_{j}^{\ast})=\check{f}_{j}\mu_{-1}
(\alpha\zeta_{j}^{\ast})e^{(\alpha-\alpha^{2})\zeta_{j}^{\ast}x
+(\alpha^{2}-\alpha)\ddot{\zeta_{j}}^{\ast}t}
\end{equation}
\begin{equation}\label{c7d}
\mu_{+3}(\alpha^{2}\zeta_{j})=\breve{f}_{j}
\mu_{-1}(\alpha^{2}\zeta_{j})e^{(\alpha^{2}-\alpha)\zeta_{j}x+
(\alpha-\alpha^{2})\ddot{\zeta_{j}}t}
\end{equation}
\begin{equation}\label{c7e}
\mu_{+3}(\zeta_{j}^{\ast})=\acute{f}_{j}\mu_{-2}
(\zeta_{j}^{\ast})e^{(\alpha-\alpha^{2})\zeta_{j}^{\ast}x+(\alpha^{2}-\alpha)\ddot{\zeta_{j}}^{\ast}t},
\end{equation}
\begin{equation}\label{c7f}
\mu_{+1}(\alpha\zeta_{j})=\grave{f}_{j}\mu_{-2} (\alpha\zeta_{j})
e^{(\alpha^{2}-\alpha)
\zeta_{j}x+(\alpha-\alpha^{2})\ddot{\zeta_{j}}t}.
\end{equation}
\end{subequations}
In order to simplify the calculation, we introduce the modified
auxiliary eigenfunctions
\begin{subequations}
\begin{equation}\label{c8a}
m_{n}(x,t,\nu)=e^{-\alpha^{2}\nu x-\alpha\nu^{-1}t}\chi_{n}(\nu),
\quad n=1, 4,
\end{equation}
\begin{equation}\label{c8b}
m_{n}(x,t,\nu)=e^{-\nu x-\nu^{-1}t}\chi_{n}(\nu), \quad n=2, 5,
\end{equation}
\begin{equation}\label{c8c}
m_{n}(x,t,\nu)=e^{-\alpha \nu x-\alpha^{2}\nu^{-1}t}\chi_{n}(\nu),
\quad n=3, 6.
\end{equation}
\end{subequations}
Let $\{\nu_{j}\}_{j=1}^{N_{2}}$ be the eigenvalues of the second type, we obtain that
\begin{subequations}
\begin{equation}\label{c9a}
\begin{aligned}
m_{2}(\nu_{j})=&\hat{c}_{j}e^{(\alpha^{2}-1)\nu_{j}x+(\alpha-1)\ddot{\nu_{j}}t}\mu_{-3}(\nu_{j}), \\
\mu_{+2}(-\nu_{j}^{\ast})=&\hat{d}_{j}e^{(\alpha-1)\nu_{j}^{\ast}x+(\alpha^{2}-1)\ddot{\nu_{j}}^{\ast}t}m_{2}(-\nu_{j}^{\ast}),
\end{aligned}
\end{equation}

\begin{equation}\label{c9b}
\begin{aligned}
m_{1}(\alpha\nu_{j}^{\ast})=&c_{j}e^{(\alpha-1)\nu_{j}^{\ast}x+(\alpha^{2}-1)\ddot{\nu_{j}}^{\ast}t}\mu_{-1}(\alpha\nu_{j}^{\ast}),\\
\mu_{+2}(-\alpha\nu_{j})=&d_{j}e^{(\alpha^{2}-1)\nu_{j}x+(\alpha-1)\ddot{\nu_{j}}t}m_{1}(-\alpha\nu_{j}),
\end{aligned}
\end{equation}

\begin{equation}\label{c9c}
\begin{aligned}
m_{3}(\alpha^{2}\nu_{j}^{\ast})=&
\check{c}_{j}e^{(\alpha-1)\nu_{j}^{\ast}x+(\alpha^{2}-1)\ddot{\nu_{j}}^{\ast}t}\mu_{-3}(\alpha^{2}\nu_{j}^{\ast}),\\
\mu_{+1}(-\alpha^{2}\nu_{j})=&\check{d}_{j}e^{(\alpha^{2}-1)\nu_{j}x+(\alpha-1)\ddot{\nu_{j}}t}m_{3}(-\alpha^{2}\nu_{j}),
\end{aligned}
\end{equation}

\begin{equation}\label{c9d}
\begin{aligned}
m_{4}(\alpha\nu_{j})=&\breve{c}_{j}
e^{(\alpha^{2}-1)\nu_{j}x+(\alpha-1)\ddot{\nu_{j}}t}\mu_{-2}(\alpha\nu_{j}),\\
\mu_{+1}(-\alpha\nu_{j}^{\ast})=&\breve{d}_{j}e^{(\alpha-1)\nu_{j}^{\ast}x+(\alpha^{2}-1)\ddot{\nu_{j}}^{\ast}t}m_{4}(-\alpha\nu_{j}^{\ast}),
\end{aligned}
\end{equation}

\begin{equation}\label{c9e}
\begin{aligned}
m_{5}(\nu_{j}^{\ast})=&\acute{c}_{j}
e^{(\alpha-1)\nu_{j}^{\ast}x+(\alpha^{2}-1)\ddot{\nu_{j}}^{\ast}t}\mu_{-2}(\nu_{j}^{\ast}),\\
\mu_{+3}(-\nu_{j})=&\acute{d}_{j}e^{(\alpha^{2}-1)\nu_{j}x+(\alpha-1)\ddot{\nu_{j}}t}m_{5}(-\nu_{j}),
\end{aligned}
\end{equation}

\begin{equation}\label{c9f}
\begin{aligned}
m_{6}(\alpha^{2}\nu_{j})=&\grave{c}_{j}
e^{(\alpha^{2}-1)\nu_{j}x+(\alpha-1)\ddot{\nu_{j}}t}\mu_{-1}(\alpha^{2}\nu_{j}),\\
\mu_{+3}(-\alpha^{2}\nu_{j}^{\ast})=&\grave{d}_{j}e^{(\alpha-1)\nu_{j}^{\ast}x+(\alpha^{2}-1)\ddot{\nu_{j}}^{\ast}t}m_{6}(-\alpha^{2}\nu_{j}^{\ast}),
\end{aligned}
\end{equation}
\end{subequations}
where $\ddot{\nu}_{j}=1/\nu_{j}$.

 The norming constants in Eqs.(3.7) and (3.9) admit the following symmetry relations
\begin{subequations}
\begin{equation}\label{c10a}
f_{j}=\grave{f}_{j}=\breve{f}_{j}, \quad
\hat{f}_{j}=\acute{f}_{j}=\check{f}_{j}, \quad f_{j}=\acute{f}^{\ast}_{j}, \quad
\end{equation}
\begin{equation}\label{c10b}
c_{j}=\check{c}_{j}=\acute{c}_{j}, \quad \hat{c}_{j}=\breve{c}_{j}=\grave{c}_{j}, \quad c^{\ast}_{j}=-\frac{\gamma^{\ast}}{\gamma}\hat{c}_{j},
\end{equation}
\begin{equation}\label{c10c}
d_{j}=\check{d}_{j}=\acute{d}_{j}, \quad
\hat{d}_{j}=\breve{d}_{j}=\grave{d}_{j},
\end{equation}
where
\begin{equation}\label{c10d}
f_{j}=-f^{\ast}_{j}\frac{\dot{r_{22}}(\nu_{j})}{\dot{s_{33}}(\nu_{j})}, \quad
d_{j}=-\frac{c^{\ast}_{j}}{s^{\ast}_{33}(-\nu^{\ast}_{j})},\quad \hat{d}_{j}=\frac{\gamma^{\ast}}{\gamma
s_{33}(-\nu^{\ast}_{j})}c_{j}.
\end{equation}
\end{subequations}
Here the dot denotes the derivative with respect to the parameter
$\nu$.
\subsection{Trace formula}
 In this section, we we consider the associated trace formula.
Assume that $s_{33}(\nu)$ has the simple zeros
$\Big\{\zeta_{j}:\zeta_{j}=-\zeta_{j}^{\ast},\zeta_{j}\in
D_{2}\Big\}_{j=1}^{N_{1}}$ and
$\Big\{\nu_{j}:\nu_{j}\neq-\nu_{j}^{\ast},\nu_{j}\in
D_{2}\Big\}_{j=1}^{N_{2}}$.

 According to the symmetry of scattering data (\ref{c3}) and (\ref{c4}), we define
\begin{subequations}
\begin{equation}\label{c11a}
\begin{aligned}
\bar{s}_{11}(\nu)=\prod_{j=1}^{N_{1}}
\frac{(\nu+\alpha^{2}\zeta_{j})(\nu+\alpha\zeta_{j}^{\ast})}
{(\nu-\alpha^{2}\zeta_{j})(\nu-\alpha\zeta_{j}^{\ast})}\prod_{j=1}^{N_{2}}\frac{(\nu+\alpha\nu_{j}^{\ast})(\nu+\alpha^{2}\nu_{j})(\nu-\nu_{j}^{\ast})(\nu-\nu_{j})}
{(\nu-\alpha\nu_{j}^{\ast})
(\nu-\alpha^{2}\nu_{j})(\nu+\nu_{j}^{\ast})(\nu+\nu_{j})}s_{11}(\nu),\\
\end{aligned}
\end{equation}
\begin{equation}\label{c11b}
\begin{aligned}
\bar{s}_{22}(\nu)=\prod_{j=1}^{N_{1}}
\frac{(\nu+\alpha\zeta_{j})(\nu+\zeta_{j}^{\ast})}
{(\nu-\alpha\zeta_{j})(\nu-\zeta_{j}^{\ast})}\prod_{j=1}^{N_{2}}\frac{
(\nu+\alpha\nu_{j})(\nu-\alpha^{2}\nu_{j})(\nu+\nu_{j}^{\ast})(\nu-\alpha^{2}\nu_{j}^{\ast})}
{(\nu-\alpha\nu_{j})
(\nu+\alpha^{2}\nu_{j})(\nu-\nu_{j}^{\ast})(\nu+\alpha^{2}\nu_{j}^{\ast})}s_{22}(\nu).\\
\end{aligned}
\end{equation}
\end{subequations}
Here, $\bar{s}_{11}(\nu)$ and $\bar{s}_{22}(\nu)$ are analytic in
$\nu\in D_{1}\cup D_{6}$ and $\nu\in D_{4}\cup D_{5}$, respectively,
whereas they have no zeros. After a few calculation, we find that
the analytic scattering coefficients of the scattering matrix take
the following form,
\begin{subequations}
\begin{equation}\label{c12a}
\begin{aligned}
s_{11}(\nu)=\prod_{j=1}^{N_{1}}
\frac{(\nu-\alpha^{2}\zeta_{j})(\nu-\alpha\zeta_{j}^{\ast})}
{(\nu+\alpha^{2}\zeta_{j})(\nu+\alpha\zeta_{j}^{\ast})}\prod_{j=1}^{N_{2}}\frac{(\nu-\alpha\nu_{j}^{\ast})
(\nu-\alpha^{2}\nu_{j})(\nu+\nu_{j}^{\ast})(\nu+\nu_{j})}
{(\nu+\alpha\nu_{j}^{\ast})(\nu+\alpha^{2}\nu_{j})(\nu-\nu_{j}^{\ast})(\nu-\nu_{j})}\\
\times\exp\bigg(\frac{1}{2\pi i}\int_{-\infty}^{+\infty}
\frac{\log(1-s_{12}(\eta)s_{12}(-\eta)-s_{13}(\eta)s_{13}(-\eta))}{\eta-\nu}d\eta\bigg),
\end{aligned}
\end{equation}
\begin{equation}\label{c12b}
\begin{aligned}
s_{22}(\nu)=\prod_{j=1}^{N_{1}}
\frac{(\nu-\alpha\zeta_{j})(\nu-\zeta_{j}^{\ast})}
{(\nu+\alpha\zeta_{j})(\nu+\zeta_{j}^{\ast})}\prod_{j=1}^{N_{2}}\frac{(\nu-\alpha\nu_{j})
(\nu+\alpha^{2}\nu_{j})(\nu-\nu_{j}^{\ast})(\nu+\alpha^{2}\nu_{j}^{\ast})}
{(\nu+\alpha\nu_{j})(\nu-\alpha^{2}\nu_{j})(\nu+\nu_{j}^{\ast})(\nu-\alpha^{2}\nu_{j}^{\ast})}\\
\times\exp\bigg(\frac{1}{2\pi i}\int_{-\infty}^{+\infty}
\frac{\log(1-s_{12}(\eta)s_{12}(-\eta)-s_{13}(\eta)s_{13}(-\eta))}{\eta-\alpha\nu}d\eta\bigg).
\end{aligned}
\end{equation}
\end{subequations}

\setcounter{equation}{0}
\section{Inverse problem}
\subsection{Riemann-Hilbert problem}
In this section, the piecewise meromorphic function
$\textbf{M}(x,t,\nu)$ in six regions for the  Tzitz\'{e}ica equation
were presented. According to the analyticities of the Jost functions and the auxiliary eigenfunctions, we define six piecewise meromorphic functions as follows
\begin{equation}\label{d1}
\begin{aligned}
\mathbf{M}_{1}(x, t, \nu)=\Bigg(\mu_{-1},\frac{\mu_{+2}}{r_{22}}, \displaystyle\frac{m_{1}}{s_{11}}\Bigg), \quad \nu\in D_{1}, \\
\mathbf{M}_{2}(x, t, \nu)=\Bigg(\displaystyle\frac{m_{2}
}{s_{33}}, \frac{\mu_{+2}}{r_{22}}, \mu_{-3}\Bigg), \nu\in D_{2}, \\
\mathbf{M}_{3}(x, t, \nu)=\Bigg(\frac{\mu_{+1}}{r_{11}}, \displaystyle\frac{m_{3} }{s_{33}}, \mu_{-3}\Bigg), \quad \nu\in D_{3}, \\
\mathbf{M}_{4}(x, t, \nu)=\Bigg(\frac{\mu_{+1}}{r_{11}}, \mu_{-2}, \displaystyle\frac{m_{4} }{s_{22}}\Bigg),\quad \nu\in D_{4}, \\
\mathbf{M}_{5}(x, t, \nu)=\Bigg(\displaystyle\frac{m_{5}}{s_{22}}, \mu_{-2}, \frac{\mu_{+3}}{r_{33}}\Bigg),\quad \nu\in D_{5}, \\
\mathbf{M}_{6}(x, t, \nu)=\Bigg(\mu_{-1},
\displaystyle\frac{m_{6}}{s_{11}},
\frac{\mu_{+3}}{r_{33}}\Bigg),\quad \nu\in D_{6}.
\end{aligned}
\end{equation}

It is readily verified that $\mathbf{M}_{n}(x,t,\nu)$ satisfy the following jump
conditions
\begin{equation}\label{d2}
\mathbf{M}_{+}(x,t,\nu)=\mathbf{M}_{-}(x,t,\nu)J(x,t,\nu), \quad
\nu\in \Sigma,
\end{equation}
where the matrices $\mathbf{M}_{+}(x,t,\nu),
\mathbf{M}_{-}(x,t,\nu)$ and $J(x,t,\nu)$ admit the following definitions
\begin{equation}\label{d3}
\mathbf{M}(x,t,\nu)=\left\{\begin{array}{ll}\mathbf{M}_{+}(x,t,\nu), \, &\nu\in D_{1}\cup D_{3}\cup D_{5},
\\ \mathbf{M}_{-}(x,t,\nu), \,&\nu\in D_{2}\cup D_{4}\cup D_{6},
\end{array}\right.
\end{equation}
\begin{equation}\label{d4}
\mathbf{J}(x,t,\nu)=\left\{\begin{array}{ll}J_{1}=
e^{\nu\hat{\sigma}x+\nu^{-1}\hat{\sigma}^{-1}t}\bigg(S_{2}^{-1}(\nu)S_{1}(\nu)\bigg),\,&\arg \nu=\frac{\pi}{3},
\\J_{2}=e^{\nu\hat{\sigma}x+\nu^{-1}\hat{\sigma}^{-1}t}\bigg(S_{2}^{-1}(\nu)S_{3}(\nu)\bigg),\,&\arg \nu=\frac{2\pi}{3},
\\J_{3}=e^{\nu\hat{\sigma}x+\nu^{-1}\hat{\sigma}^{-1}t}\bigg(S_{4}^{-1}(\nu)S_{3}(\nu)\bigg),\,&\arg \nu=\pi,
\\J_{4}=e^{\nu\hat{\sigma}x+\nu^{-1}\hat{\sigma}^{-1}t}\bigg(S_{4}^{-1}(\nu)S_{5}(\nu)\bigg),\,&\arg \nu=\frac{4\pi}{3},
\\J_{5}=e^{\nu\hat{\sigma}x+\nu^{-1}\hat{\sigma}^{-1}t}\bigg(S_{6}^{-1}(\nu)S_{5}(\nu)\bigg),\,&\arg \nu=\frac{5\pi}{3},
\\J_{6}=e^{\nu\hat{\sigma}x+\nu^{-1}\hat{\sigma}^{-1}t}\bigg(S_{6}^{-1}(\nu)S_{1}(\nu)\bigg),\,&\arg
\nu=0.
\end{array}\right.
\end{equation}
We note that the jump conditions (\ref{d2}) can be rewritten as
\begin{equation}\label{d5}
\begin{aligned}
&\mathbf{M}_{+}(x,t,\nu)-\mathbf{M}_{-}(x,t,\nu)\\
&=\mathbf{M}_{-}(x,t,\nu)\bigg(e^{\nu\sigma x+\nu^{-1}\sigma^{-1}t}\mathbf{\hat{J}}(x,t,\nu)e^{-\nu\sigma x-\nu^{-1}\sigma^{-1}t}\bigg), \quad \nu\in \Sigma,
\end{aligned}
\end{equation}
where the jump matrices $\hat{J}_{1}, \ldots, \hat{J}_{6}$  can be
derived as
\begin{equation}\label{d6}
\begin{scriptsize}
\begin{aligned}
\mathbf{\hat{J}}_{1}=
\left(\begin{matrix}
0&0&-\rho_{1}(\nu)\\
0&0&0\\
\rho_{2}(\nu)&0&-\rho_{1}(\nu)\rho_{2}(\nu)
\end{matrix}\right), \nu \in \Sigma_{1},
\quad
\mathbf{\hat{J}}_{2}=
\left(\begin{matrix}
-\rho^{\ast}_{1}(-\nu^{\ast})\rho^{\ast}_{2}(-\nu^{\ast})&-\rho^{\ast}_{2}(-\nu^{\ast})&0 \\
\rho^{\ast}_{1}(-\nu^{\ast})&0&0\\
0&0&0
\end{matrix}\right), \nu\in\Sigma_{2},\\
\mathbf{\hat{J}}_{3}=
\left(\begin{matrix}
0&0&0\\
0&-\rho^{\ast}_{3}(\nu^{\ast})\rho_{3}(\nu)&\rho_{3}(\nu) \\
0&-\rho^{\ast}_{3}(\nu^{\ast})&0
\end{matrix}\right), \nu \in \Sigma_{3},
\quad
\mathbf{\hat{J}}_{4}=
\left(\begin{matrix}
0&0& \rho_{2}(-\nu)\\
0&0&0\\
-\rho_{1}(-\nu)&0&-\rho_{1}(-\nu)\rho_{2}(-\nu)
\end{matrix}\right), \nu \in \Sigma_{4},\\
\mathbf{\hat{J}}_{5}=
\left(\begin{matrix}
-\rho^{\ast}_{1}(\nu^{\ast})\rho^{\ast}_{2}(\nu^{\ast})&
\rho^{\ast}_{1}(\nu^{\ast})&0\\
-\rho^{\ast}_{2}(\nu^{\ast})&0&0\\
0&0&0
\end{matrix}\right), \nu\in \Sigma_{5},
\quad
\mathbf{\hat{J}}_{6}=
\left(\begin{matrix}
0&0&0 \\
0&-\rho_{3}(-\nu)\rho^{\ast}_{3}(-\nu^{\ast})& -\rho^{\ast}_{3}(-\nu^{\ast})\\
0&\rho_{3}(-\nu)&0
\end{matrix}\right), \nu\in\Sigma_{6}.\\
\end{aligned}
\end{scriptsize}
\end{equation}

To find the normalization condition of the Riemann-Hilbert problem, we note that, from Eq. (2.14), asymptotic behaviors ${\mu_{\pm}(x,t,\nu)}$ take the following form
\begin{equation}\label{d7}
 \mu_{\pm}(x,t,\nu)=\left(\begin{matrix}
1&\alpha^{2} &\alpha\\
1&\alpha&\alpha^{2}\\
1&1&1
\end{matrix}\right)+\frac{1}{\nu}\left(\begin{matrix}
\eta-\frac{2}{3}u_{x}&\alpha\eta-\frac{2}{3}\alpha u_{x}&\alpha^{2}\eta-\frac{2}{3}\alpha^{2}u_{x}\\
\eta+\frac{1}{3}u_{x}&\eta+\frac{1}{3}u_{x}&\eta+\frac{1}{3}u_{x}\\
\eta+\frac{1}{3}u_{x}&\alpha^{2}\eta+\frac{1}{3}\alpha^{2}
u_{x}&\alpha\eta+\frac{1}{3}\alpha u_{x}
\end{matrix}\right)+o(\frac{1}{\nu^{2}}), \nu\longrightarrow \infty,
\end{equation}
where $\eta=\frac{1}{3}\int_{\pm\infty}^{x}u_{\xi}^{2} d\xi$, which implies that
 $$\mathbf{M}_{\pm}(x,t,\nu)\rightarrow A, \quad \nu\rightarrow \infty .$$

\subsection{The case of no poles}
In this case, matrix functions $\mathbf{M}_{\pm}(x,t,\nu)$ are the analytic in their respective
domains.
Applying the Cauchy operator $\frac{1}{2\pi i}\int_{\Sigma}[\frac{f}{\zeta-\nu}]d\zeta$ to the equation (\ref{d5}), we have the following integral representation
\begin{equation}\label{d8}
\mathbf{M}_{\pm}(x,t,\nu)=A+\frac{1}{2\pi
i}\int_{\Sigma_{2}\cup\Sigma_{4}\cup\Sigma_{6}}\frac{\bigg(M_{n}\bar{J}_{n}\bigg)(\zeta)}{\zeta-\nu}d\zeta
+\frac{1}{2\pi
i}\int_{\Sigma_{1}\cup\Sigma_{3}\cup\Sigma_{5}}\frac{\bigg(M_{n+1}\bar{J}_{n}\bigg)(\zeta)}{\zeta-\nu}d\zeta,
\end{equation}
where
$$\zeta \in \Sigma_{n}, \quad \bar{J}_{n}=e^{\zeta\sigma
x+\zeta^{-1}\sigma^{-1}t}\hat{J}_{n}e^{-\zeta\sigma
x-\zeta^{-1}\sigma^{-1}t}.$$
\subsection{The case of poles}
In this case, we assume that $s_{33}(\nu)$ has the simple zeros
$\{\zeta_{j}:\zeta_{j}=-\zeta_{j}^{\ast}\}_{j=1}^{N_{1}}$ and
$\{\nu_{j}:\nu_{j}\neq-\nu_{j}^{\ast}\}_{j=1}^{N_{2}}$ in $D_{2}$.
 The meromorphic matrices defined in (\ref{d1})
satisfy the following residue conditions
\begin{subequations}
\begin{equation}\label{d9a}
\mathbf{M}_{1,\alpha\zeta_{j}^{\ast}}(x,t,\nu)=\check{F}_{j}\bigg(\textbf{0},\mu_{-1}(\alpha\zeta_{j}^{\ast}),\textbf{0}\bigg), \mathbf{M}_{2,\zeta_{j}}(x,t,\nu)=F_{j}\bigg(\textbf{0},\mu_{-3}(\zeta_{j}),\textbf{0}\bigg),
\end{equation}
\begin{equation}\label{d9b}
\mathbf{M}_{3,\alpha^{2}\zeta_{j}^{\ast}}(x,t,\nu)=\hat{F}_{j}
\bigg(\mu_{-3}(\alpha^{2}\zeta_{j}^{\ast}),\textbf{0},\textbf{0}\bigg), \mathbf{M}_{4,\alpha\zeta_{j}}(x,t,\nu)=\grave{F}_{j}
\bigg(\mu_{-2}(\alpha\zeta_{j}),\textbf{0},\textbf{0}\bigg),
\end{equation}
\begin{equation}\label{d9c}
\mathbf{M}_{5,\zeta_{j}^{\ast}}(x,t,\nu)=\acute{F}_{j}
\bigg(\textbf{0},\textbf{0},\mu_{-2}(\zeta_{j}^{\ast})\bigg), \mathbf{M}_{6,\alpha^{2}\zeta_{j}}(x,t,\alpha^{2}\zeta_{j})=
\breve{F}_{j}\bigg(\textbf{0},\textbf{0},\mu_{-1}(\alpha^{2}\zeta_{j})\bigg),
\end{equation}
\end{subequations}
\begin{subequations}
\begin{equation}\label{d10a}
\mathbf{M}_{1,-\alpha\nu_{j}}(x,t,\nu)=D_{j}\bigg(\textbf{0},\frac{m_{1}}{s_{11}}(-\alpha\nu_{j}),\textbf{0}\bigg), \mathbf{M}_{1,\alpha\nu_{j}^{\ast}}(x,t,\nu)=C_{j}\bigg(\textbf{0},\textbf{0},\mu_{-1}(\alpha\nu_{j}^{\ast})\bigg),
\end{equation}
\begin{equation}\label{d10b}
\mathbf{M}_{2,-\nu_{j}^{\ast}}(x,t,\nu)=\hat{D}_{j}\bigg(\textbf{0},\frac{m_{2}}{s_{33}}(-\nu_{j}^{\ast}),\textbf{0}\bigg), \mathbf{M}_{2,\nu_{j}}(x,t,\nu)=\hat{C}_{j}\bigg(\mu_{-3}(\nu_{j}),\textbf{0},\textbf{0}\bigg),
\end{equation}
\begin{equation}\label{d10c}
\mathbf{M}_{3,-\alpha^{2}\nu_{j}}(x,t,\nu)=
\check{D}_{j}\bigg(\frac{m_{3}}{s_{33}}(-\alpha^{2}\nu_{j}),\textbf{0},\textbf{0}\bigg), \mathbf{M}_{3,\alpha^{2}\nu_{j}^{\ast}}(x,t,\nu)=\check{C}_{j}\bigg(\textbf{0},\mu_{-3}(\alpha^{2}\nu_{j}^{\ast}),\textbf{0}\bigg),
\end{equation}
\begin{equation}\label{d10d}
\mathbf{M}_{4,-\alpha\nu_{j}^{\ast}}(x,t,\nu)=
\breve{D}_{j}\bigg(\frac{m_{4}}{s_{22}}(-\alpha\nu_{j}^{\ast}),\textbf{0},\textbf{0}\bigg), \mathbf{M}_{4,\alpha\nu_{j}}(x,t,\nu)=\breve{C}_{j}\bigg(\textbf{0},\textbf{0},\mu_{-2}(\alpha\nu_{j})\bigg),
\end{equation}
\begin{equation}\label{d10e}
\mathbf{M}_{5,-\nu_{j}}(x,t,\nu)=\acute{D}_{j}\bigg(\textbf{0},\textbf{0},\frac{m_{5}}{s_{22}}(-\nu_{j})\bigg), \mathbf{M}_{5,\nu_{j}^{\ast}}(x,t,\nu)=
\acute{C}_{j}\bigg(\mu_{-2}(\nu_{j}^{\ast}),\textbf{0},\textbf{0}\bigg),
\end{equation}
\begin{equation}\label{d10f}
\mathbf{M}_{6,-\alpha^{2}\nu_{j}^{\ast}}(x,t,\nu)=\grave{D}_{j}\bigg
(\textbf{0},\textbf{0},\frac{m_{6}}{s_{11}}(-\alpha^{2}\nu_{j}^{\ast})\bigg), \mathbf{M}_{6,\alpha^{2}\nu_{j}}(x,t,\alpha^{2}\nu_{j})=
\grave{C}_{j}\bigg(\textbf{0},\mu_{-1}(\alpha^{2}\nu_{j}),\textbf{0}\bigg),
\end{equation}
\end{subequations}
where
\begin{subequations}
\begin{equation}\label{d11a}
\check{F}_{j}(x,t,\nu)=\frac{\check{f}_{j}}{\dot{r_{22}}
(\alpha\zeta_{j}^{\ast})}e^{(\alpha-\alpha^{2})\zeta_{j}^{\ast}x+(\alpha^{2}-\alpha)\ddot{\zeta_{j}}^{\ast}t}
, F_{j}(x,t,\nu)= \frac{f_{j}}{\dot{r_{22}}(\zeta_{j})}
e^{(\alpha^{2}-\alpha)\zeta_{j}x+(\alpha-\alpha^{2})\ddot{\zeta_{j}}t},
\end{equation}
\begin{equation}\label{d11b}
\hat{F}_{j}(x,t,\nu)=\frac{\hat{f}_{j}}{\dot{r_{11}}
(\alpha^{2}\zeta_{j}^{\ast})}
e^{(\alpha-\alpha^{2})\zeta_{j}^{\ast}x+(\alpha^{2}-\alpha)\ddot{\zeta_{j}}^{\ast}t},
\grave{F}_{j}(x,t,\nu)= \frac{\grave{f}_{j}}{\dot{r_{11}}
(\alpha\zeta_{j})}
e^{(\alpha^{2}-\alpha)\zeta_{j}x+(\alpha-\alpha^{2})\ddot{\zeta_{j}}t},
\end{equation}
\begin{equation}\label{d11c}
\acute{F}_{j}(x,t,\nu)=\frac{\acute{f}_{j}}{\dot{r_{33}}
(\zeta_{j}^{\ast})}
e^{(\alpha-\alpha^{2})\zeta_{j}^{\ast}x+(\alpha^{2}-\alpha)\ddot{\zeta_{j}}^{\ast}t},
\breve{F}_{j}(x,t,\nu)= \frac{\breve{f}_{j}}{\dot{r_{33}}
(\alpha^{2}\zeta_{j})}
e^{(\alpha^{2}-\alpha)\zeta_{j}x+(\alpha-\alpha^{2})\ddot{\zeta_{j}}t},
\end{equation}
\end{subequations}
\begin{subequations}
\begin{equation}\label{d12a}
D_{j}(x,t)=\frac{d_{j}s_{11}(-\alpha\nu_{j})}{\dot{r_{22}}(-\alpha\nu_{j})}
e^{(\alpha^{2}-1)\nu_{j}x+(\alpha-1)\ddot{\nu_{j}}t},
C_{j}(x,t)=\frac{c_{j}}{\dot{s_{11}}
(\alpha\nu_{j}^{\ast})}e^{(\alpha-1)\nu_{j}^{\ast}x
+(\alpha^{2}-1)\ddot{\nu_{j}}^{\ast}t},
\end{equation}
\begin{equation}\label{d12b}
\hat{D}_{j}(x,t)=\frac{\hat{d}_{j}s_{33}(-\nu^{\ast}_{j})}{\dot{r_{22}}(-\nu_{j}^{\ast})}
e^{(\alpha-1)\nu_{j}^{\ast}x+(\alpha^{2}-1)\ddot{\nu_{j}}^{\ast}t},
\hat{C}_{j}(x,t)=\frac{\hat{c}_{j}}{\dot{s_{33}}(\nu_{j})}
e^{(\alpha^{2}-1)\nu_{j}x+(\alpha-1)\ddot{\nu_{j}}t},
\end{equation}
\begin{equation}\label{d12c}
\check{D}_{j}(x,t)=\frac{\check{d}_{j}s_{33}(-\alpha^{2}\nu_{j})}{\dot{r_{11}}(-\alpha^{2}\nu_{j})}
e^{(\alpha^{2}-1)\nu_{j}x+(\alpha-1)\ddot{\nu_{j}}t},
\check{C}_{j}(x,t)=
\frac{\check{c}_{j}}{\dot{s_{33}}(\alpha^{2}\nu_{j}^{\ast})}
e^{(\alpha-1)\nu_{j}^{\ast}x+(\alpha^{2}-1)\ddot{\nu_{j}}^{\ast}t},
\end{equation}
\begin{equation}\label{d12d}
\breve{D}_{j}(x,t)=\frac{\breve{d}_{j}s_{22}(-\alpha\nu^{\ast}_{j})}{\dot{r_{11}}(-\alpha\nu_{j}^{\ast})}
e^{(\alpha-1)\nu_{j}^{\ast}x+(\alpha^{2}-1)\ddot{\nu_{j}}^{\ast}t},
\breve{C}_{j}(x,t)=\frac{\breve{c}_{j}}
{\dot{s_{22}}(\alpha\nu_{j})}e^{(\alpha^{2}-1)\nu_{j}x+(\alpha-1)\ddot{\nu_{j}}t},
\end{equation}
\begin{equation}\label{d12e}
\acute{D}_{j}(x,t)=\frac{\acute{d}_{j}(x,t)s_{22}(-\nu_{j})}{\dot{r_{33}}(-\nu_{j})}
e^{(\alpha^{2}-1)\nu_{j}x+(\alpha-1)\ddot{\nu_{j}}t},
\acute{C}_{j}(x,t)=\frac{\acute{c}_{j}}{\dot{s_{22}}(\nu_{j}^{\ast})}
e^{(\alpha-1)\nu_{j}^{\ast}x+(\alpha^{2}-1)\ddot{\nu_{j}}^{\ast}t},
\end{equation}
\begin{equation}\label{d12f}
\grave{D}_{j}(x,t)=\frac{\grave{d}_{j}s_{11}(-\alpha^{2}\nu_{j}^{\ast})}{\dot{r_{33}}
(-\alpha^{2}\nu_{j}^{\ast})}e^{(\alpha-1)\nu_{j}^{\ast}x+(\alpha^{2}-1)\ddot{\nu_{j}}^{\ast}t},
\grave{C}_{j}(x,t)=\frac{\grave{c}_{j}}{\dot{s_{11}}
(\alpha^{2}\nu_{j})}e^{(\alpha^{2}-1)\nu_{j}x+(\alpha-1)\ddot{\nu_{j}}t}.
\end{equation}
\end{subequations}
It is remarked that
$$\begin{aligned}
&\acute{F}_{j}(x,t,\nu)=F_{j}^{\ast}(x,t,\nu), \check{F}_{j}(x,t,\nu)
=\alpha F_{j}^{\ast}(x,t,\nu), \grave{F}_{j}(x,t,\nu)=\alpha F_{j}(x,t,\nu),\\
&\hat{F}_{j}(x,t,\nu)=\alpha^{2}F_{j}^{\ast}(x,t,\nu), \breve{F}_{j}(x,t,\nu)=\alpha^{2}F_{j}(x,t,\nu),\\
&\grave{D}_{j}(x,0,\nu)=\alpha^{2}\hat{D}_{j}(x,t,\nu)=\alpha\breve{D}_{j}(x,t,\nu), {D}_{j}(x,t,\nu)=\alpha
\acute{D}_{j}(x,t,\nu)=\alpha^{2}\check{D}_{j}(x,t,\nu),\\
\end{aligned}$$
and
\begin{eqnarray*}
&&{C}_{j}(x,t,\nu)=\alpha\acute{C}_{j}(x,t,\nu)=\alpha^{2}\check{C}_{j}(x,t,\nu),
\grave{C}_{j}(x,t,\nu)=\alpha^{2}\hat{C}_{j}(x,t,\nu)=\alpha\breve{C}_{j}(x,t,\nu),\\
&&\acute{D}_{j}(x,t,\nu)=(\acute{C}_{j}(x,t,\nu))^{\ast},
\hat{C}_{j}(x,t,\nu)=(\hat{D}_{j}(x,t,\nu))^{\ast},\\
&&\hat{D}_{j}(x,t,\nu)=\acute{C}_{j} (x,t,\nu),\hat{C}_{j}(x,t,\nu)=(\acute{C}_{j}(x,t,\nu))^{\ast}.
\end{eqnarray*}

The key to solving the RH problem of (4.2) is to convert it into a
mixed system of algebraic-integral equations. We similarly apply
$\frac{1}{2\pi i}\int_{\Sigma}[\frac{f}{\zeta-\nu}]d\zeta$ to both
sides of the jump condition (\ref{d5}).  Taking into account the
symmetries properties of the involved functions, 
we can easily obtain
\begin{equation}\label{d13}
\begin{aligned}
m_{1}^{\pm}(x, t, \nu)=&\left(\begin{array}{c}
1\\[-2pt]
1\\[-2pt]
1
\end{array}\right)
+\sum_{j=1}^{N_{1}}\left\{\frac{\hat{F}_{j}\mu_{-3}(\alpha^{2}\zeta_{j}^{\ast})}{\nu-\alpha^{2}\zeta_{j}^{\ast}}
+\frac{\grave{F}_{j}\mu_{-2}(\alpha\zeta_{j})}{\nu-\alpha\zeta_{j}}\right\}
+\sum_{j=1}^{N_{2}}\left\{\frac{\hat{C}_{j}\mu_{-3}(\nu_{j})}{\nu-\nu_{j}}\right.
\\ &\left.
+\frac{\check{D}_{j}}{\nu+\alpha^{2}\nu_{j}}\frac{m_{3}}{s_{33}}(-\alpha^{2}\nu_{j})
+\frac{\acute{C}_{j}\mu_{-2}(\nu_{j}^{\ast})}{\nu-\nu_{j}^{\ast}}
+\frac{\breve{D}_{j}}{\nu+\alpha\nu_{j}^{\ast}}\frac{m_{4}}{s_{22}}(-\alpha\nu_{j}^{\ast})\right\}\\
&+\frac{1}{2 \pi i}\int_{\Sigma}\frac{(M^{-}\bar{J}_{n})_{1}(\zeta)}{\zeta-z}d\zeta,
\end{aligned}
\end{equation}
\begin{equation}\label{d14}
\begin{aligned}
m_{2}^{+}(x, t, \nu)=&\left(\begin{array}{c}
\alpha^{2}\\[-1pt]
\alpha\\[-1pt]
1
\end{array}\right)+\sum_{j=1}^{N_{1}}\left\{\frac{\check{F}_{j}\mu_{-1}(\alpha\zeta_{j}^{\ast})}{\nu-\alpha\zeta_{j}^{\ast}}
+\frac{F_{j}\mu_{-3}(\zeta_{j})}{\nu-\zeta_{j}}\right\}
+\sum_{j=1}^{N_{2}}\left\{\displaystyle\frac{D_{j}\frac{m_{1}}{s_{11}}(-\alpha\nu_{j})}{\nu+\alpha\nu_{j}}\right. \\   &\left. +\frac{\hat{D}_{j}}{\nu+\nu_{j}^{\ast}}\frac{m_{2}}{s_{33}}(-\nu_{j}^{\ast})
+\frac{\check{C}_{j}\mu_{-3}(\alpha^{2}\nu_{j}^{\ast})}{\nu-\alpha^{2}\nu_{j}^{\ast}}
+\frac{\grave{C}_{j}\mu_{-1}(\alpha^{2}\nu_{j})}{\nu-\alpha^{2}\nu_{j}}\right\}\\
&+\frac{1}{2 \pi
i}\int_{\Sigma}\displaystyle\frac{(M^{-}\bar{J}_{n})_{2}(\zeta)}{\zeta-z}d\zeta,
\end{aligned}
\end{equation}
\begin{equation}\label{d15}
\begin{aligned}
m_{2}^{-}(x, t, \nu)=&\left(\begin{array}{c}
\alpha^{2}\\[-1pt]
\alpha\\[-1pt]
1
\end{array}\right)+\sum_{j=1}^{N_{1}}\left\{\displaystyle\frac{\check{F}_{j}\mu_{-1}(\alpha\zeta_{j}^{\ast})}{\nu-\alpha\zeta_{j}^{\ast}}
+\displaystyle\frac{F_{j}\mu_{-3}(\zeta_{j})}{\nu-\zeta_{j}}\right\}
+\sum_{j=1}^{N_{2}}\left\{\displaystyle\frac{D_{j}\frac{m_{1}}{s_{11}}(-\alpha\nu_{j}) }{\nu+\alpha\nu_{j}}\right. \\  &\left. +\displaystyle\frac{\hat{D}_{j}}{\nu+\nu_{j}^{\ast}}\frac{m_{2}}{s_{33}}(-\nu_{j}^{\ast})
+\displaystyle\frac{\check{C}_{j}\mu_{-3}(\alpha^{2}\nu_{j}^{\ast})}{\nu-\alpha^{2}\nu_{j}^{\ast}}
+\displaystyle\frac{\grave{C}_{j}\mu_{-1}(\alpha^{2}\nu_{j})}{\nu-\alpha^{2}\nu_{j}}\right\}\\
&+\frac{1}{2 \pi i}\int_{\Sigma}\displaystyle\frac{(M^{-}\bar{J}_{n})_{2}(\zeta)}{\zeta-z}d\zeta,
\end{aligned}
\end{equation}
\begin{equation}\label{d16}
\begin{aligned}
m_{3}^{+}(x, t, \nu)=&\left(\begin{array}{c}
\alpha\\[-1pt]
\alpha^{2}\\[-1pt]
1
\end{array}\right)+\sum_{j=1}^{N_{1}}\left\{\displaystyle\frac{\acute{F}_{j}\mu_{-2}(\zeta_{j}^{\ast})}{\nu-\zeta_{j}^{\ast}}
+\displaystyle\frac{\breve{F}_{j}\mu_{-1}(\alpha^{2}\zeta_{j})}{\nu-\alpha^{2}\zeta_{j}}\right\}
+\sum_{j=1}^{N_{2}}\left\{\frac{C_{j}\mu_{-1}(\alpha\nu_{j}^{\ast})}{\nu-\alpha\nu_{j}^{\ast}}\right. \\ &\left. +\displaystyle\frac{\breve{C}_{j}\mu_{-2}(\alpha\nu_{j})}{\nu-\alpha\nu_{j}}
+\displaystyle\frac{\acute{D}_{j}}{\nu+\nu_{j}}\frac{m_{5}}{s_{22}}(-\nu_{j})
+\displaystyle\frac{\grave{D}_{j}}{\nu+\alpha^{2}\nu_{j}^{\ast}}\frac{m_{6}}{s_{11}}(-\alpha^{2}\nu_{j}^{\ast})\right\}\\
&+\frac{1}{2 \pi i}\int_{\Sigma}\displaystyle\frac{(M^{-}\bar{J}_{n})_{3}(\zeta)}{\zeta-z}d\zeta ,
\end{aligned}
\end{equation}
\begin{equation}\label{d17}
\begin{aligned}
m_{3}^{-}(x, t, \nu)=&\left(\begin{array}{c}
\alpha\\[-1pt]
\alpha^{2}\\[-1pt]
1
\end{array}\right)+\sum_{j=1}^{N_{1}}\left\{\displaystyle\frac{\acute{F}_{j}
\mu_{-2}(\zeta_{j}^{\ast})}{\nu-\zeta_{j}^{\ast}}
+\displaystyle\frac{\breve{F}_{j}\mu_{-1}(\alpha^{2}\zeta_{j})}{\nu-\alpha^{2}\zeta_{j}}\right\}
+\sum_{j=1}^{N_{2}}\left\{\displaystyle\frac{C_{j}\mu_{-1}(\alpha\nu_{j}^{\ast})}{\nu-\alpha\nu_{j}^{\ast}}\right. \\   &\left. +\displaystyle\frac{\breve{C}_{j}\mu_{-2}(\alpha\nu_{j})}{\nu-\alpha\nu_{j}}
+\displaystyle\frac{\acute{D}_{j}}{\nu+\nu_{j}}\frac{m_{5}}{s_{22}}(-\nu_{j})
+\displaystyle\frac{\grave{D}_{j}}{\nu+\alpha^{2}\nu_{j}^{\ast}}\frac{m_{6}}{s_{11}}(-\alpha^{2}\nu_{j}^{\ast})\right\}\\
&+\frac{1}{2 \pi i}\int_{\Sigma}\displaystyle\frac{(M^{-}\bar{J}_{n})_{3}(\zeta)}{\zeta-z}d\zeta. 
\end{aligned}
\end{equation}

To finish the potential reconstruction about the scattering data, we consider the expansion of (\ref{d13}) at $\nu\to\infty$, and compare it with (\ref{d7}). And the $O(\nu^{-1})$ terms imply the representation of the potential matrix about the eigenfunctions at certain eigenvalues, which can be reduced by taking $\nu=-\alpha^{2}\nu_{j}, \nu_{j}^{\ast}$ in (\ref{d14}), $\nu=\alpha\zeta_{j}$ in (\ref{d15}), $\nu=\alpha^{2}\zeta_{j}^{\ast}$ in (\ref{d16}) and $\nu=\nu_{j},-\alpha\nu_{j}^{\ast}$ in (\ref{d17}). Then the potential can be reconstructed as
\begin{equation}\label{d18}
\begin{aligned}
 U_{x}=\sum_{j=1}^{N_{1}}\Bigg\{\hat{F}_{j}\bigg(\mu_{-23}-\mu_{-13}\bigg)(\alpha^{2}\zeta_{j}^{\ast})+\grave{F}_{j}
\bigg(\mu_{-22}-\mu_{-12}\bigg)(\alpha\zeta_{j})\Bigg\}\\
+\sum_{j=1}^{N_{2}}\Bigg\{\hat{C}_{j}\bigg(\mu_{-23}-\mu_{-13}\bigg)(\nu_{j})
+\check{D}_{j}\bigg(\frac{m_{23}}{s_{33}}-\frac{m_{13}}{s_{33}}\bigg)(-\alpha^{2}\nu_{j})\\
+\acute{C}_{j}\bigg(\mu_{-22}-\mu_{-12}\bigg)(\nu_{j}^{\ast})+\breve{D}_{j}\bigg(\frac{m_{24}}{s_{22}}-\frac{m_{14}}{s_{22}}\bigg)(-\alpha\nu_{j}^{\ast})\Bigg\}\\
-\frac{1}{2 \pi i}\int_{\Sigma}\Big((M^{-}\bar{J}_{n})_{21}-(M^{-}\bar{J}_{n})_{11}\Big)(\zeta)d\zeta .
\end{aligned}
\end{equation}

\setcounter{equation}{0}
\section{Explicit solutions }
In the reflectionless case, it is convenient to write (\ref{d17}) as fractional equation
$$U_{x}=\frac{\det(G^{\prime}+b^{\ast}Y^{T})}{\det G^{\prime}}-\frac{\det(G+bY^{T})}{\det G},$$
where
\begin{eqnarray*}
G=I-F,\quad G^{\prime}=I-F^{\prime},\quad
b=(b_{1}, \ldots, b_{2N_{1}+4N_{2}})^{T},\quad Y=(y_{1}, \ldots, y_{2N_{1}+4N_{2}})^{T},
\end{eqnarray*}
and  $F_{ij}, b_{j}, y_{j}$ are given by (A.1)-(A.2), (A.4)-(A.9)in the appendix.\\
(1)$I$-explicit solution.\\
Assuming $N_{1}=1$ and $N_{2}=0$, and setting
$\zeta_{1}=ie^{s},s\in \mathbb{R},c_{0},d_{0}\in\mathbb{R}$,
we then find the following explicit solution
$$u(x,t,\nu)=\partial^{-1}\Bigg\{-\displaystyle
\frac{6e^{s}\sinh(c)\sin(d-\frac{\pi}{3})}
{\sinh^{2}(c)+2\cos(d+\frac{2\pi}{3})\cosh(c)+2\cos(2d-\frac{2\pi}{3})}\Bigg \},$$
where
$$c=c_{0}+\sqrt{3}(e^{s}x+e^{-s}t),d=d_{0}.$$
(2)$II$-explicit solution.\\
Choosing $N_{1}=0$ and $N_{2}=1$, and supposing
$\nu_{1}=e^{\epsilon_{1}+i\eta_{1}}, \epsilon_{0}, \eta_{0},
\epsilon_{1}, \eta_{1}\in\mathbb{R}, $
we get another explicit solution for the Eq. (\ref{a1}) as
\begin{equation}\label{e1}
u(x,t,\nu)=\partial^{-1}\Bigg
\{3\sqrt{3}e^{\epsilon_{1}}\displaystyle \frac{A} {B}\Bigg \},
\end{equation}
where
\begin{equation*}
\begin{split}
A=&-\omega\sin2b+e^{2\xi-2\tau}\tan\eta_{1}\omega\sinh(2\tau+2a)\\
&-\frac{3}{8}\omega e^{2\xi-\tau+\epsilon_{0}}[\sin(s+\gamma)\cos b \sinh(\tau+a)
-\sin b\cos(s+\gamma)\cosh(\tau+a)]\\
&+\frac{1}{2}e^{6\xi-3\tau}[\sin
b \cos s \cosh(3a+3\tau)+\sin s\cos b\sinh(3a+3\tau)]\\
&-\frac{1}{2}e^{2\xi-\tau}[\sin s\cos b \sinh(a+\tau)+\sin3b \cos s
\cosh(a+\tau)],
\end{split}
\end{equation*}
\begin{equation*}\label{e2}
\begin{split}
B=&\Bigg[\cos
b-e^{2\xi-\tau}\cosh(a+\tau)\Bigg]^{2}\Bigg\{e^{4\xi-2\tau}\cosh^{2}(a+\tau)+\cos^{2}
b-3\frac{\cos^{2}(\eta_{1}-\frac{\pi}{3})}{\cos\eta_{1}\cos(\eta_{1}+\frac{\pi}{3})}\Bigg\},
\end{split}
\end{equation*}
with
\begin{eqnarray*}
&&a=\epsilon_{0}+e^{\epsilon_{1}}
\Bigg(\frac{\sqrt{3}}{2}\sin\eta_{1}-
\frac{3}{2}\cos\eta_{1}\Bigg)x-e^{-\epsilon_{1}}\Bigg(\frac{3}{2}\cos\eta_{1}-\frac{\sqrt{3}}{2}\sin\eta_{1}\Bigg)t,\\
&&b=\eta_{0}+e^{\epsilon_{1}}
\Bigg(\frac{3}{2}\sin\eta_{1}+
\frac{\sqrt{3}}{2}\cos\eta_{1}\Bigg)x-e^{-\epsilon_{1}}\Bigg(\frac{\sqrt{3}}{2}\cos\eta_{1}+\frac{3}{2}\sin\eta_{1}\Bigg)t,\\
&&s=\eta_{1}+\frac{2}{3}\pi, \quad\tau=
\frac{1}{2}\ln\Bigg\{\frac{\tan\eta_{1}}{\tan(\eta_{1}+\frac{\pi}{3})}\Bigg\},\quad
 \xi=\frac{1}{2}\ln\Bigg\{\frac{\tan\eta_{1}}{\tan(\eta_{1}-\frac{\pi}{3})}\Bigg\},\\
&&\gamma=-\frac{2\pi i}{3}-\frac{i}{2}
\ln\Bigg\{\displaystyle\frac{e^{i(4\eta_{1}-\frac{2\pi}{3})}+3e^{-4i\eta_{1}}+2e^{\frac{2\pi
i}{3}}}{2+e^{-i(4\eta_{1}+\frac{2\pi}{3})}+3e^{i(4\eta_{1}+\frac{2\pi}{3})}}\Bigg\},\\
&&\omega=\frac{\cos^{3}(\eta_{1}-\frac{\pi}{3})}{\cos\eta_{1}\cos(\eta_{1}+\frac{\pi}{3})}
.
\end{eqnarray*}
\appendix
\setcounter{equation}{0}
\section*{Appendix}
\section{Reflectionless potentials}
Through calculating, we can give the coefficients $b_{j}(x,t)$ and
$y_{j}(x,t)$ as follows
\begin{equation}\label{f1}
b_{j}(x,t,\nu)=\left\{\begin{array}{ll}\alpha^{2},j=1,\ldots,N_{1},
\\ \alpha,j=N_{1}+1,\ldots,2N_{1},
\\ \alpha,j=2N_{1}+1,\ldots,2N_{1}+N_{2},
\\ \alpha^{2},j=2N_{1}+N_{2}+1,\ldots,2N_{1}+2N_{2},
\\ \alpha^{2},j=2N_{1}+2N_{2}+1,\ldots,2N_{1}+3N_{2},
\\ \alpha,j=2N_{1}+3N_{2}+1,\ldots,2N_{1}+4N_{2},
\end{array}\right.
\end{equation}
\begin{equation}\label{f2}
y_{j}(x,t,\nu)=\left\{\begin{array}{ll}\grave{F}_{j},j=1,\ldots,N_{1},
\\ \hat{F}_{j-N_{1}},j=N_{1}+1,\ldots,2N_{1},
\\ \hat{C}_{j-2N_{1}},j=2N_{1}+1,\ldots,2N_{1}+N_{2},
\\ \acute{C}_{j-2N_{1}-N_{2}},j=2N_{1}+N_{2}+1,\ldots,2N_{1}+2N_{2},
\\ \check{D}_{j-2N_{1}-2N_{2}},j=2N_{1}+2N_{2}+1,\ldots,2N_{1}+3N_{2},
\\ \breve{D}_{j-2N_{1}-3N_{2}},j=2N_{1}+3N_{2}+1,\ldots,2N_{1}+4N_{2}.
\end{array}\right.
\end{equation}
For convenience, these definitions are given
\begin{subequations}
\begin{equation}\label{f3a}
F_{n}^{(1)}(x,t,\nu)=F_{n}(x,t,\nu)\frac{\alpha^{2}}{\nu-\zeta_{n}},F_{n}^{(2)}(x,t,\nu)=
\check{F}_{n}(x,t,\nu)\frac{\alpha^{2}}{\nu-\alpha\zeta_{n}^{\ast}},
\end{equation}
\begin{equation}\label{f3b}
F_{n}^{(3)}(x,t,\nu)=\breve{F}_{n}(x,t,\nu)\frac{\alpha}{\nu-\alpha^{2}\zeta_{n}},
F_{n}^{(4)}(x,t,\nu)=\acute{F}_{n}(x,t,\nu)\frac{\alpha}{\nu-\zeta_{n}^{\ast}},
\end{equation}
\begin{equation}\label{f3c}
C_{n}^{(1)}(x,t,\nu)=\grave{C}_{n}(x,t,\nu)\frac{\alpha^{2}}{\nu-\alpha^{2}\nu_{n}},
C_{n}^{(2)}(x,t,\nu)=\check{C}_{n}(x,t,\nu)\frac{\alpha^{2}}{\nu-\alpha^{2}\nu_{n}^{\ast}},
\end{equation}
\begin{equation}\label{f3d}
C_{n}^{(3)}(x,t,\nu)=\breve{C}_{n}(x,t,\nu)\frac{\alpha}{\nu-\alpha\nu_{n}},
C_{n}^{(4)}(x,t,\nu)=C_{n}(x,t,\nu)\frac{\alpha}{\nu-\alpha\nu_{n}^{\ast}},
\end{equation}
\begin{equation}\label{f3e}
D_{n}^{(1)}(x,t,\nu)=D_{n}(x,t,\nu)\frac{\alpha^{2}}{\nu+\alpha\nu_{n}},
D_{n}^{(2)}(x,t,\nu)=\hat{D}_{n}(x,t,\nu)\frac{\alpha^{2}}{\nu+\nu_{n}^{\ast}},
\end{equation}
\begin{equation}\label{f3f}
D_{n}^{(3)}(x,t,\nu)=\acute{D}_{n}(x,t,\nu)\frac{\alpha}{\nu+\nu_{n}},
D_{n}^{(4)}(x,t,\nu)=\grave{D}_{n}(x,t,\nu)\frac{\alpha}{\nu+\alpha^{2}\nu_{n}^{\ast}}.
\end{equation}
\end{subequations}
Additionally, these terms $F_{ij}$ can be derived from (A.3)
\begin{subequations}
\begin{equation}\label{f4a}
F_{ij}(x,t,\nu)=F_{j}^{(1)}(\alpha\zeta_{i})~(i, j=1, \ldots, N_{1}),
\end{equation}
For $i=1,\ldots, N_{1}, j=N_{1}+1,\ldots, 2N_{1}$,
\begin{equation}\label{f4b}
F_{ij}(x,t,\nu)=F_{j-N_{1}}^{(2)}(\alpha\zeta_{i}).
\end{equation}
For $i=1,\ldots, N_{1}, j=2N_{1}+1,\ldots, 2N_{1}+N_{2}$,
\begin{equation}\label{f4c}
F_{ij}(x,t,\nu)=C_{j-2N_{1}}^{(1)}(\alpha\zeta_{i}).
\end{equation}
For $i=1,\ldots, N_{1}, j=2N_{1}+N_{2}+1,\ldots, 2N_{1}+2N_{2}$,
\begin{equation}\label{f4d}
F_{ij}(x,t,\nu)=C_{j-2N_{1}-N_{2}}^{(2)}(\alpha\zeta_{i}).
\end{equation}
For $i=1,\ldots, N_{1}, j=2N_{1}+2N_{2}+1,\ldots, 2N_{1}+3N_{2}$,
\begin{equation}\label{f4e}
F_{ij}(x,t,\nu)=D_{j-2N_{1}-2N_{2}}^{(1)}(\alpha\zeta_{i}).
\end{equation}
For $i=1,\ldots, N_{1}, j=2N_{1}+3N_{2}+1,\ldots, 2N_{1}+4N_{2}$,
\begin{equation}\label{f4f}
F_{ij}(x,t,\nu)=D_{j-2N_{1}-3N_{2}}^{(2)}(\alpha\zeta_{i}).
\end{equation}
\end{subequations}
For $i=N_{1}+1,\ldots, 2N_{1}, j=1,\ldots, N_{1}$,
\begin{subequations}
\begin{equation}\label{f5a}
F_{ij}(x,t,\nu)=F_{j}^{(3)}(\alpha^{2}\zeta_{i-N_{1}}^{\ast}).
\end{equation}
For $i=N_{1}+1,\ldots, 2N_{1}, j=N_{1}+1,\ldots, 2N_{1}$,
\begin{equation}\label{f5b}
F_{ij}(x,t,\nu)=F_{j-N_{1}}^{(4)}(\alpha^{2}\zeta_{i-N_{1}}^{\ast}).
 \end{equation}
 For $i=N_{1}+1,\ldots, 2N_{1}, j=2N_{1}+1,\ldots, 2N_{1}+N_{2}$,
\begin{equation}\label{f5c}
F_{ij}(x,t,\nu)=C_{j-2N_{1}}^{(3)}(\alpha^{2}\zeta_{i-N_{1}}^{\ast}).
\end{equation}
For $i=N_{1}+1,\ldots, 2N_{1}, j=2N_{1}+N_{2}+1,\ldots,
2N_{1}+2N_{2}$,
\begin{equation}\label{f5d}
F_{ij}(x,t,\nu)=C_{j-2N_{1}-N_{2}}^{(4)}(\alpha^{2}\zeta_{i-N_{1}}^{\ast}).
\end{equation}
For $i=N_{1}+1,\ldots, 2N_{1}, j=2N_{1}+2N_{2}+1,\ldots,
2N_{1}+3N_{2}$,
\begin{equation}\label{f5e}
F_{ij}(x,t,\nu)=D_{j-2N_{1}-2N_{2}}^{(3)}(\alpha^{2}\zeta_{i-N_{1}}^{\ast}).
\end{equation}
For $i=N_{1}+1,\ldots, 2N_{1}, j=2N_{1}+3N_{2}+1,\ldots,
2N_{1}+4N_{2}$,
\begin{equation}\label{f5f}
F_{ij}(x,t,\nu)=D_{j-2N_{1}-3N_{2}}^{(4)}(\alpha^{2}\zeta_{i-N_{1}}^{\ast}).
\end{equation}
\end{subequations}
For $i=2N_{1}+1,\ldots, 2N_{1}+N_{2}, j=1,\ldots, N_{1}$,
\begin{subequations}
\begin{equation}\label{f6a}
F_{ij}(x,t,\nu)=F_{j}^{(3)}(\nu_{i-2N_{1}}).
\end{equation}
For $i=2N_{1}+1,\ldots, 2N_{1}+N_{2}, j=N_{1}+1,\ldots, 2N_{1}$,
\begin{equation}\label{f6b}
F_{ij}(x,t,\nu)=F_{j-N_{1}}^{(4)}(\nu_{i-2N_{1}}).
 \end{equation}
 For $i=2N_{1}+1,\ldots, 2N_{1}+N_{2}, j=2N_{1}+1,\ldots, 2N_{1}+N_{2}$,
\begin{equation}\label{f6c}
F_{ij}(x,t,\nu)=C_{j-2N_{1}}^{(3)}(\nu_{i-2N_{1}}).
\end{equation}
For $i=2N_{1}+1,\ldots, 2N_{1}+N_{2}, j=2N_{1}+N_{2}+1,\ldots,
2N_{1}+2N_{2}$,
\begin{equation}\label{f6d}
F_{ij}(x,t,\nu)=C_{j-2N_{1}-N_{2}}^{(4)}(\nu_{i-2N_{1}}).
\end{equation}
For $i=2N_{1}+1,\ldots, 2N_{1}+N_{2}, j=2N_{1}+2N_{2}+1,\ldots,
2N_{1}+3N_{2}$,
\begin{equation}\label{f6e}
F_{ij}(x,t,\nu)=D_{j-2N_{1}-2N_{2}}^{(3)}(\nu_{i-2N_{1}}).
\end{equation}
For $i=2N_{1}+1,\ldots, 2N_{1}+N_{2}, j=2N_{1}+3N_{2}+1,\ldots,
2N_{1}+4N_{2}$,
\begin{equation}\label{f6f}
F_{ij}(x,t,\nu)=D_{j-2N_{1}-3N_{2}}^{(4)}(\nu_{i-2N_{1}}).
\end{equation}
\end{subequations}
For $i=2N_{1}+N_{2}+1,\ldots, 2N_{1}+2N_{2}, j=1,\ldots, N_{1}$,
\begin{subequations}
\begin{equation}\label{f7a}
F_{ij}(x,t,\nu)=F_{j}^{(1)}(\nu_{i-2N_{1}-N_{2}}^{\ast}).
\end{equation}
For $i=2N_{1}+N_{2}+1,\ldots, 2N_{1}+2N_{2}, j=N_{1}+1,\ldots,
2N_{1}$,
\begin{equation}\label{f7b}
F_{ij}(x,t,\nu)=F_{j-N_{1}}^{(2)}(\nu_{i-2N_{1}-N_{2}}^{\ast}).
\end{equation}
For $i=2N_{1}+N_{2}+1,\ldots, 2N_{1}+2N_{2}, j=2N_{1}+1,\ldots,
2N_{1}+N_{2}$,
\begin{equation}\label{f7c}
F_{ij}(x,t,\nu)=C_{j-2N_{1}}^{(1)}(\nu_{i-2N_{1}-N_{2}}^{\ast}).
\end{equation}
For $i=2N_{1}+N_{2}+1,\ldots, 2N_{1}+2N_{2},
j=2N_{1}+N_{2}+1,\ldots, 2N_{1}+2N_{2}$,
\begin{equation}\label{f7d}
F_{ij}(x,t,\nu)=C_{j-2N_{1}-N_{2}}^{(2)}(\nu_{i-2N_{1}-N_{2}}^{\ast}).
\end{equation}
For $i=2N_{1}+N_{2}+1,\ldots, 2N_{1}+2N_{2},
j=2N_{1}+2N_{2}+1,\ldots, 2N_{1}+3N_{2}$,
\begin{equation}\label{f7e}
F_{ij}(x,t,\nu)=D_{j-2N_{1}-2N_{2}}^{(1)}(\nu_{i-2N_{1}-N_{2}}^{\ast}).
\end{equation}
For $i=2N_{1}+N_{2}+1,\ldots, 2N_{1}+2N_{2},
j=2N_{1}+3N_{2}+1,\ldots, 2N_{1}+4N_{2}$,
\begin{equation}\label{f7f}
F_{ij}(x,t,\nu)=D_{j-2N_{1}-3N_{2}}^{(2)}(\nu_{i-2N_{1}-N_{2}}^{\ast}).
\end{equation}
\end{subequations}
For $i=2N_{1}+2N_{2}+1,\ldots, 2N_{1}+3N_{2}, j=1,\ldots, N_{1}$,
\begin{subequations}
\begin{equation}\label{f8a}
F_{ij}(x,t,\nu)=F_{j}^{(1)}(-\alpha^{2}\nu_{i-2N_{1}-2N_{2}}).
\end{equation}
For $i=2N_{1}+2N_{2}+1,\ldots, 2N_{1}+3N_{2}, j=N_{1}+1,\ldots,
2N_{1}$,
\begin{equation}\label{f8b}
F_{ij}(x,t,\nu)=F_{j-N_{1}}^{(2)}(-\alpha^{2}\nu_{i-2N_{1}-2N_{2}}).
 \end{equation}
For $i=2N_{1}+2N_{2}+1,\ldots, 2N_{1}+3N_{2}, j=2N_{1}+1,\ldots,
2N_{1}+N_{2}$,
\begin{equation}\label{f8c}
F_{ij}(x,t,\nu)=C_{j-2N_{1}}^{(1)}(-\alpha^{2}\nu_{i-2N_{1}-2N_{2}}).
\end{equation}
For $i=2N_{1}+2N_{2}+1,\ldots, 2N_{1}+3N_{2},
j=2N_{1}+N_{2}+1,\ldots, 2N_{1}+2N_{2}$,
\begin{equation}\label{f8d}
F_{ij}(x,t,\nu)=C_{j-2N_{1}-N_{2}}^{(2)}(-\alpha^{2}\nu_{i-2N_{1}-2N_{2}}).
\end{equation}
For $i=2N_{1}+2N_{2}+1,\ldots, 2N_{1}+3N_{2},
j=2N_{1}+2N_{2}+1,\ldots, 2N_{1}+3N_{2}$,
\begin{equation}\label{f8e}
F_{ij}(x,t,\nu)=D_{j-2N_{1}-2N_{2}}^{(1)}(-\alpha^{2}\nu_{i-2N_{1}-2N_{2}}).
\end{equation}
For $i=2N_{1}+2N_{2}+1,\ldots, 2N_{1}+3N_{2},
j=2N_{1}+3N_{2}+1,\ldots, 2N_{1}+4N_{2}$,
\begin{equation}\label{f8f}
F_{ij}(x,t,\nu)=D_{j-2N_{1}-3N_{2}}^{(2)}(-\alpha^{2}\nu_{i-2N_{1}-2N_{2}}).
\end{equation}
\end{subequations}
For $i=2N_{1}+3N_{2}+1,\ldots, 2N_{1}+4N_{2}, j=1,\ldots, N_{1}$,
\begin{subequations}
\begin{equation}\label{f9a}
F_{ij}(x,t,\nu)=F_{j}^{(3)}(-\alpha\nu_{i-2N_{1}-3N_{2}}^{\ast}).
\end{equation}
For $i=2N_{1}+3N_{2}+1,\ldots, 2N_{1}+4N_{2}, j=N_{1}+1,\ldots,
2N_{1}$,
\begin{equation}\label{f9b}
F_{ij}(x,t,\nu)=F_{j-N_{1}}^{(4)}(-\alpha\nu_{i-2N_{1}-3N_{2}}^{\ast}).
 \end{equation}
For $i=2N_{1}+3N_{2}+1,\ldots, 2N_{1}+4N_{2}, j=2N_{1}+1,\ldots,
2N_{1}+N_{2}$,
\begin{equation}\label{f9c}
F_{ij}(x,t,\nu)=C_{j-2N_{1}}^{(3)}(-\alpha\nu_{i-2N_{1}-3N_{2}}^{\ast}).
\end{equation}
For $i=2N_{1}+3N_{2}+1,\ldots, 2N_{1}+4N_{2},
j=2N_{1}+N_{2}+1,\ldots, 2N_{1}+2N_{2}$,
\begin{equation}\label{f9d}
F_{ij}(x,t,\nu)=C_{j-2N_{1}-N_{2}}^{(4)}(-\alpha\nu_{i-2N_{1}-3N_{2}}^{\ast}).
\end{equation}
For $i=2N_{1}+3N_{2}+1,\ldots, 2N_{1}+4N_{2},
j=2N_{1}+2N_{2}+1,\ldots, 2N_{1}+3N_{2}$,
\begin{equation}\label{f9e}
F_{ij}(x,t,\nu)=D_{j-2N_{1}-2N_{2}}^{(3)}(-\alpha\nu_{i-2N_{1}-3N_{2}}^{\ast}).
\end{equation}
For $i=2N_{1}+3N_{2}+1,\ldots, 2N_{1}+4N_{2},
j=2N_{1}+3N_{2}+1,\ldots, 2N_{1}+4N_{2}$,
\begin{equation}\label{f9f}
F_{ij}(x,t,\nu)=D_{j-2N_{1}-3N_{2}}^{(4)}(-\alpha\nu_{i-2N_{1}-3N_{2}}^{\ast}).
\end{equation}
\end{subequations}
For $i, j=1,\ldots, N_{1}$,
\begin{subequations}
\begin{equation}\label{f10a}
F_{ij}^{\prime}(x,t,\nu)=\alpha^{2}F_{j}^{(1)}(\alpha\zeta_{i}).
\end{equation}
For $i=1,\ldots, N_{1}, j=N_{1}+1,\ldots, 2N_{1}$,
\begin{equation}\label{f10b}
F_{ij}^{\prime}(x,t,\nu)=\alpha^{2}F_{j-N_{1}}^{(2)}(\alpha\zeta_{i}).
 \end{equation}
 For $i=1,\ldots, N_{1}, j=2N_{1}+1,\ldots, 2N_{1}+N_{2}$,
\begin{equation}\label{f10c}
F_{ij}^{\prime}(x,t,\nu)=\alpha^{2}C_{j-2N_{1}}^{(1)}(\alpha\zeta_{i}).
\end{equation}
For $i=1,\ldots, N_{1}, j=2N_{1}+N_{2}+1,\ldots, 2N_{1}+2N_{2}$,
\begin{equation}\label{f10d}
F_{ij}^{\prime}(x,t,\nu)=\alpha^{2}C_{j-2N_{1}-N_{2}}^{(2)}(\alpha\zeta_{i}).
\end{equation}
For $i=1,\ldots, N_{1}, j=2N_{1}+2N_{2}+1,\ldots, 2N_{1}+3N_{2}$,
\begin{equation}\label{f10e}
F_{ij}^{\prime}(x,t,\nu)=\alpha^{2}D_{j-2N_{1}-2N_{2}}^{(1)}(\alpha\zeta_{i}).
\end{equation}
For $i=1,\ldots, N_{1}, j=2N_{1}+3N_{2}+1,\ldots, 2N_{1}+4N_{2}$,
\begin{equation}\label{f10f}
F_{ij}^{\prime}(x,t,\nu)=\alpha^{2}D_{j-2N_{1}-3N_{2}}^{(2)}(\alpha\zeta_{i}).
\end{equation}
\end{subequations}
For $i=N_{1}+1,\ldots, 2N_{1}, j=1,\ldots, N_{1}$,
\begin{subequations}
\begin{equation}\label{f11a}
F_{ij}^{\prime}(x,t,\nu)=\alpha F_{j}^{(3)}(\alpha^{2}\zeta_{i-N_{1}}^{\ast}).
\end{equation}
For $i=N_{1}+1,\ldots, 2N_{1}, j=N_{1}+1,\ldots, 2N_{1}$,
\begin{equation}\label{f11b}
F_{ij}^{\prime}(x,t,\nu)=\alpha F_{j-N_{1}}^{(4)}(\alpha^{2}\zeta_{i-N_{1}}^{\ast}).
 \end{equation}
 For $i=N_{1}+1,\ldots, 2N_{1}, j=2N_{1}+1,\ldots, 2N_{1}+N_{2}$,
\begin{equation}\label{f11c}
F_{ij}^{\prime}(x,t,\nu)=\alpha C_{j-2N_{1}}^{(3)}(\alpha^{2}\zeta_{i-N_{1}}^{\ast}).
\end{equation}
For $i=N_{1}+1,\ldots, 2N_{1}, j=2N_{1}+N_{2}+1,\ldots,
2N_{1}+2N_{2}$,
\begin{equation}\label{f11d}
F_{ij}^{\prime}(x,t,\nu)=\alpha C_{j-2N_{1}-N_{2}}^{(4)}(\alpha^{2}\zeta_{i-N_{1}}^{\ast}).
\end{equation}
For $i=N_{1}+1,\ldots, 2N_{1}, j=2N_{1}+2N_{2}+1,\ldots,
2N_{1}+3N_{2}$,
\begin{equation}\label{f11e}
F_{ij}^{\prime}(x,t,\nu)=\alpha D_{j-2N_{1}-2N_{2}}^{(3)}(\alpha^{2}\zeta_{i-N_{1}}^{\ast}).
\end{equation}
For $i=N_{1}+1,\ldots, 2N_{1}, j=2N_{1}+3N_{2}+1,\ldots,
2N_{1}+4N_{2}$,
\begin{equation}\label{f11f}
F_{ij}^{\prime}(x,t,\nu)=\alpha D_{j-2N_{1}-3N_{2}}^{(4)}(\alpha^{2}\zeta_{i-N_{1}}^{\ast}).
\end{equation}
\end{subequations}
For $i=2N_{1}+1,\ldots, 2N_{1}+N_{2}, j=1,\ldots, N_{1}$,
\begin{subequations}
\begin{equation}\label{f12a}
F_{ij}^{\prime}(x,t,\nu)=\alpha F_{j}^{(3)}(\nu_{i-2N_{1}}).
\end{equation}
For $i=2N_{1}+1,\ldots, 2N_{1}+N_{2}, j=N_{1}+1,\ldots, 2N_{1}$,
\begin{equation}\label{f12b}
F_{ij}^{\prime}(x,t,\nu)=\alpha F_{j-N_{1}}^{(4)}(\nu_{i-2N_{1}}).
 \end{equation}
 For $i=2N_{1}+1,\ldots, 2N_{1}+N_{2}, j=2N_{1}+1,\ldots, 2N_{1}+N_{2}$,
\begin{equation}\label{f12c}
F_{ij}^{\prime}(x,t,\nu)=\alpha C_{j-2N_{1}}^{(3)}(\nu_{i-2N_{1}}).
\end{equation}
For $i=2N_{1}+1,\ldots, 2N_{1}+N_{2}, j=2N_{1}+N_{2}+1,\ldots,
2N_{1}+2N_{2}$,
\begin{equation}\label{f12d}
F_{ij}^{\prime}(x,t,\nu)=\alpha C_{j-2N_{1}-N_{2}}^{(4)}(\nu_{i-2N_{1}}).
\end{equation}
For $i=2N_{1}+1,\ldots, 2N_{1}+N_{2}, j=2N_{1}+2N_{2}+1,\ldots,
2N_{1}+3N_{2}$,
\begin{equation}\label{f12e}
F_{ij}^{\prime}(x,t,\nu)=\alpha D_{j-2N_{1}-2N_{2}}^{(3)}(\nu_{i-2N_{1}}).
\end{equation}
For $i=2N_{1}+1,\ldots, 2N_{1}+N_{2}, j=2N_{1}+3N_{2}+1,\ldots,
2N_{1}+4N_{2}$,
\begin{equation}\label{f12f}
F_{ij}^{\prime}(x,t,\nu)=\alpha D_{j-2N_{1}-3N_{2}}^{(4)}(\nu_{i-2N_{1}}).
\end{equation}
\end{subequations}
For $i=2N_{1}+N_{2}+1,\ldots, 2N_{1}+2N_{2}, j=1,\ldots, N_{1}$,
\begin{subequations}
\begin{equation}\label{f13a}
F_{ij}^{\prime}(x,t,\nu)=\alpha^{2}F_{j}^{(1)}(\nu_{i-2N_{1}-N_{2}}^{\ast}).
\end{equation}
For $i=2N_{1}+N_{2}+1,\ldots, 2N_{1}+2N_{2}, j=N_{1}+1,\ldots,
2N_{1}$,
\begin{equation}\label{f13b}
F_{ij}^{\prime}(x,t,\nu)=\alpha^{2}F_{j-N_{1}}^{(2)}(\nu_{i-2N_{1}-N_{2}}^{\ast}).
 \end{equation}
For $i=2N_{1}+N_{2}+1,\ldots, 2N_{1}+2N_{2}, j=2N_{1}+1,\ldots,
2N_{1}+N_{2}$,
\begin{equation}\label{f13c}
F_{ij}^{\prime}(x,t,\nu)=\alpha^{2}C_{j-2N_{1}}^{(1)}(\nu_{i-2N_{1}-N_{2}}^{\ast}).
\end{equation}
For $i=2N_{1}+N_{2}+1,\ldots, 2N_{1}+2N_{2},
j=2N_{1}+N_{2}+1,\ldots, 2N_{1}+2N_{2}$,
\begin{equation}\label{f13d}
F_{ij}^{\prime}(x,t,\nu)=\alpha^{2}C_{j-2N_{1}-N_{2}}^{(2)}(\nu_{i-2N_{1}-N_{2}}^{\ast}).
\end{equation}
For $i=2N_{1}+N_{2}+1,\ldots, 2N_{1}+2N_{2},
j=2N_{1}+2N_{2}+1,\ldots, 2N_{1}+3N_{2}$,
\begin{equation}\label{f13e}
F_{ij}^{\prime}(x,t,\nu)=\alpha^{2}D_{j-2N_{1}-2N_{2}}^{(1)}(\nu_{i-2N_{1}-N_{2}}^{\ast}).
\end{equation}
For $i=2N_{1}+N_{2}+1,\ldots, 2N_{1}+2N_{2},
j=2N_{1}+3N_{2}+1,\ldots, 2N_{1}+4N_{2}$,
\begin{equation}\label{f13f}
F_{ij}^{\prime}(x,t,\nu)=\alpha^{2}D_{j-2N_{1}-3N_{2}}^{(2)}(\nu_{i-2N_{1}-N_{2}}^{\ast}).
\end{equation}
\end{subequations}
For $i=2N_{1}+2N_{2}+1,\ldots, 2N_{1}+3N_{2}, j=1, \ldots, N_{1}$,
\begin{subequations}
\begin{equation}\label{f14a}
F_{ij}^{\prime}(x,t,\nu)=\alpha^{2}F_{j}^{(1)}(-\alpha^{2}\nu_{i-2N_{1}-2N_{2}}).
\end{equation}
For $i=2N_{1}+2N_{2}+1,\ldots, 2N_{1}+3N_{2}, j=N_{1}+1,\ldots,
2N_{1}$,
\begin{equation}\label{f14b}
F_{ij}^{\prime}(x,t,\nu)=\alpha^{2}F_{j-N_{1}}^{(2)}(-\alpha^{2}\nu_{i-2N_{1}-2N_{2}}).
 \end{equation}
For $i=2N_{1}+2N_{2}+1,\ldots, 2N_{1}+3N_{2}, j=2N_{1}+1,\ldots,
2N_{1}+N_{2}$,
\begin{equation}\label{f14c}
F_{ij}^{\prime}(x,t,\nu)=\alpha^{2}C_{j-2N_{1}}^{(1)}(-\alpha^{2}\nu_{i-2N_{1}-2N_{2}}).
\end{equation}
For $i=2N_{1}+2N_{2}+1,\ldots, 2N_{1}+3N_{2},
j=2N_{1}+N_{2}+1,\ldots, 2N_{1}+2N_{2}$,
\begin{equation}\label{f14d}
F_{ij}^{\prime}(x,t,\nu)=\alpha^{2}C_{j-2N_{1}-N_{2}}^{(2)}(-\alpha^{2}\nu_{i-2N_{1}-2N_{2}}).
\end{equation}
For $i=2N_{1}+2N_{2}+1,\ldots, 2N_{1}+3N_{2},
j=2N_{1}+2N_{2}+1,\ldots, 2N_{1}+3N_{2}$,
\begin{equation}\label{f14e}
F_{ij}^{\prime}(x,t,\nu)=\alpha^{2}D_{j-2N_{1}-2N_{2}}^{(1)}(-\alpha^{2}\nu_{i-2N_{1}-2N_{2}}).
\end{equation}
For $i=2N_{1}+2N_{2}+1,\ldots, 2N_{1}+3N_{2},
j=2N_{1}+3N_{2}+1,\ldots, 2N_{1}+4N_{2}$,
\begin{equation}\label{f14f}
F_{ij}^{\prime}(x,t,\nu)=\alpha^{2}D_{j-2N_{1}-3N_{2}}^{(2)}(-\alpha^{2}\nu_{i-2N_{1}-2N_{2}}).
\end{equation}
\end{subequations}
For $i=2N_{1}+3N_{2}+1,\ldots, 2N_{1}+4N_{2}, j=1,\ldots, N_{1}$,
\begin{subequations}
\begin{equation}\label{f15a}
F_{ij}^{\prime}(x,t,\nu)=\alpha F_{j}^{(3)}(-\alpha\nu_{i-2N_{1}-3N_{2}}^{\ast}).
\end{equation}
For $i=2N_{1}+3N_{2}+1,\ldots, 2N_{1}+4N_{2}, j=N_{1}+1,\ldots,
2N_{1}$,
\begin{equation}\label{f15b}
F_{ij}^{\prime}(x,t,\nu)=\alpha F_{j-N_{1}}^{(4)}(-\alpha\nu_{i-2N_{1}-3N_{2}}^{\ast}).
 \end{equation}
For $i=2N_{1}+3N_{2}+1,\ldots, 2N_{1}+4N_{2}, j=2N_{1}+1,\ldots,
2N_{1}+N_{2}$,
\begin{equation}\label{f15c}
F_{ij}^{\prime}(x,t,\nu)=\alpha C_{j-2N_{1}}^{(3)}(-\alpha\nu_{i-2N_{1}-3N_{2}}^{\ast}).
\end{equation}
For $i=2N_{1}+3N_{2}+1,\ldots, 2N_{1}+4N_{2},
j=2N_{1}+N_{2}+1,\ldots, 2N_{1}+2N_{2}$,
\begin{equation}\label{f15d}
F_{ij}^{\prime}(x,t,\nu)=\alpha C_{j-2N_{1}-N_{2}}^{(4)}(-\alpha\nu_{i-2N_{1}-3N_{2}}^{\ast}).
\end{equation}
For $i=2N_{1}+3N_{2}+1,\ldots, 2N_{1}+4N_{2},
j=2N_{1}+2N_{2}+1,\ldots, 2N_{1}+3N_{2}$,
\begin{equation}\label{f15e}
F_{ij}^{\prime}(x,t,\nu)=\alpha D_{j-2N_{1}-2N_{2}}^{(3)}(-\alpha\nu_{i-2N_{1}-3N_{2}}^{\ast}).
\end{equation}
Finally, for $i=2N_{1}+3N_{2}+1,\ldots, 2N_{1}+4N_{2},
j=2N_{1}+3N_{2}+1, \ldots, 2N_{1}+4N_{2}$,
\begin{equation}\label{f15f}
F_{ij}^{\prime}(x,t,\nu)=\alpha D_{j-2N_{1}-3N_{2}}^{(4)}(-\alpha\nu_{i-2N_{1}-3N_{2}}^{\ast}).
\end{equation}
\end{subequations}
\section*{Acknowledgments}
The projects 11871440, 11931017 and 11801144 are all supported by
the National Natural Science Foundation of China.

\end{document}